\documentclass[fleqn,usenatbib,useAMS]{mnras}
\usepackage{multicol}        
\usepackage{bm}		
\usepackage{pdflscape}	

\usepackage{graphicx,epsf,epsfig}
\usepackage{inputenc} 
\usepackage{amsmath}
\usepackage{amsfonts}
\usepackage{mathrsfs}
\usepackage{amssymb}
\usepackage{amscd}
\usepackage{dcolumn}
\usepackage{bm}
\usepackage{natbib}
\usepackage{url}
\usepackage{xspace}

\usepackage[normalem]{ulem}
\usepackage{appendix}
\usepackage{algorithm}
\usepackage{algorithmic}

\usepackage{xcolor}

\def\be{\begin{equation}}
\def\ee{\end{equation}}
\def\bea{\begin{eqnarray}}
\def\eea{\end{eqnarray}}

\def\prl{Phys. Rev. Lett.}

\def\prc{Phys. Rev. C}
\def\prd{Phys. Rev. D}

\def\mnras{Mon. Not. Roy. Astr. Soc.}
\def\apj{Astrophys. J.}
\def\apjl{Astrophys. J. Lett.}

\def\nphysa{Nucl. Phys.}

\def\jcap{J.C.A.P.}

\title{Quasi-periodic oscillations in rotating and deformed spacetimes} 

\author[K.~Boshkayev, et. al.]{K.~Boshkayev,$^{1,2,3}$\thanks{kuantay@mail.ru} 
T.~Konysbayev,$^{1,2}$\thanks{talgar\_777@mail.ru} Ye.~Kurmanov,$^{1,2,4}$\thanks{kurmanov.yergali@kaznu.kz} M.~Muccino,$^{2,5}$\thanks{marco.muccino@lnf.infn.it} \newauthor  and H.~ Quevedo,$^{2,6,7}$\thanks{quevedo@nucleares.unam.mx}\\
$^1$National Nanotechnology Laboratory of Open Type,  Almaty 050040, Kazakhstan.\\
$^2$Al-Farabi Kazakh National University, Al-Farabi ave. 71, 050040 Almaty , Kazakhstan.\\
$^3$International Information Technology University, Manas st. 34/1, 050040 Almaty, Kazakhstan.\\
$^4${International Engineering Technological University, 93G/5 Al-Farabi avenue, 050060 Almaty, Kazakhstan.}\\
$^5$Universit\`a di Camerino, Via Madonna delle Carceri 9, 62032 Camerino, Italy\\
$^6$Instituto de Ciencias Nucleares, Universidad Nacional Aut\`onoma de M\`exico, Mexico.\\
$^7$Dipartimento di Fisica and Icra, Universit\`a di Roma “La Sapienza”, Roma, Italy.}

\begin{document}

\date{\today}
\maketitle

\begin{abstract}
Quasi-periodic oscillation (QPOs) analysis is important for understanding the dynamical behavior of many astrophysical objects during transient events such as gamma-ray bursts, solar flares, magnetar flares, and fast radio bursts. In this paper, we analyze QPO data in low-mass X-ray binary (LMXB) systems, using the Lense-Thirring, Kerr, and approximate Zipoy-Voorhees metrics. We demonstrate that the inclusion of spin and quadrupole parameters modifies the well-established results for the fundamental frequencies in the Schwarzschild spacetime. We interpret the QPO data within the framework of the standard relativistic precession model, allowing us to infer the values of the mass, spin, and quadrupole parameters of neutron stars in LMXBs. We explore recent QPO data sets from eight distinct LMXBs, assess their optimal parameters, and compare our findings with results in the existing literature. Finally, we discuss the astrophysical implications of our findings.
\end{abstract}


\section{Introduction}

Recently, general relativity has been primarily tested through experiments within the solar system and binary pulsar observations \citep{2014LRR....17....4W}. The constraints on potential deviations from Einstein's theory of gravity in weak gravitational fields are currently quite stringent. The testing of general relativity in the realm of strong gravity represents a new frontier. There are two possible approaches to testing the nature of black holes (BHs) and neutron stars (NSs). 1) One can examine the properties of electromagnetic radiation emitted by accreting gas near these celestial bodies \citep{2017RvMP...89b5001B, 2016CQGra..33f4001B, 2016CQGra..33k3001J}. Presently, the two primary techniques involve thin-disk thermal spectroscopy \citep{2011ApJ...731..121B, 2012ApJ...761..174B, 2014ApJ...797...78K} and the analysis of reflected spectroscopy (iron line) \citep{2013ApJ...773...57J, 2013PhRvD..87b3007B, 2013PhRvD..88f4022B, 2015ApJ...811..130J}, but new methods may become available in the future. 2) One can study the gravitational waves emitted by BHs \citep{2013LRR....16....9Y, 2016CQGra..33e4001Y}. Until very recently, this was purely speculative, but it has now become a reality \citep{2016PhRvL.116f1102A}.

Quasi-periodic oscillations (QPOs) manifest in the variable characteristics of numerous low-mass X-ray binaries (LMXBs), including those containing NSs. Within this category of oscillations, there are kilohertz (or high-frequency) QPOs, typically characterized by frequencies denoted as $f_L$ (lower) and $f_U$ (upper), which typically fall within the range of 50-1300 Hz. This frequency range is of a similar order of magnitude as the frequencies associated with orbital motion in the vicinity of a compact object. Consequently, most models explaining kilohertz QPOs incorporate the concept of orbital motion occurring in the inner regions of the accretion disk.

The most suitable environment for testing the effects of strong gravity lies in the vicinity of astrophysical compact objects like BHs and NSs. According to the principles of general relativity, the spacetime surrounding spherically symmetric objects is exactly described by the Schwarzschild solution \citep{1973grav.book.....M}. Furthermore, for slowly rotating objects, the Lense-Thirring solution remains applicable to both BHs and NSs. However, the situation significantly alters when rapid rotation comes into play. In such scenarios, the spacetime around rapidly spinning BHs is characterized by the Kerr solution. In contrast, the spacetime around fast-rotating NSs requires more intricate solutions, which must take into account the quadrupolar deformation of these stars. Furthermore, there is no unique solution for NSs due to the presence of the mass quadrupole moment.

Initial deviations from Kerr's metric are quickly dissipated through the emission of gravitational waves \citep{1972PhRvD...5.2419P}. The equilibrium electric charge is entirely negligible for macroscopic objects \citep{2009JCAP...09..013B}. The presence of the accretion disk is irrelevant, and its impact on the background metric can be easily ignored \citep{2014PhRvD..89l7302B}. Correspondingly, the accretion disk can be modelled in terms of test particles moving around compact objects.

Numerous exact and approximate solutions to the Einstein field equations are available in the literature \citep{solutions, bqr12}. In this context, we primarily concentrate on one particular exterior solution: the Zipoy-Voorhees metric, which is also referred to in the literature as the $\gamma$-metric, $\delta$-metric, and $q$-metric \citep{mala04, mala05, Mashhoon2018, quev11, TMD}. This metric characterizes the gravitational field of static and deformed astrophysical objects and represents the simplest extension of the Schwarzschild metric, incorporating the quadrupole parameter \citep{quev12}.

In the present paper, we envisage the orbital (circular) and epicyclic motion of test particles around a compact object. In addition, we analyze the fundamental frequencies of the different metrics (models) and QPOs observed in the eight particular source such as Cir X-1, GX 5-1, GX 17+2, GX 340+0, Sco X1, 4U1608-52, 4U1728-34, 4U0614+091 \citep{2023PhRvD.108d4063B, 2023arXiv230715003B, 2022arXiv221210186B}. The purpose of this work is to study QPOs in the Lense-Thirring, Kerr and approximate Zipoy-Voorhees solutions when the deformation of the source is small. Throughout this article, the signature of the line element  is adopted as ($-$, $+$ ,$+$ ,$+$) and geometric units are used ($G=c=1$). 
However, for expressions that have astrophysical significance, we explicitly use physical units.

This work is organized as follows: in Sections  \ref{MetricLT},   \ref{sez2b}, and \ref{metric}, we review the main properties and the QPOs of the Lense-Thirring, Kerr, and Zipoy-Voorhees metrics, respectively.  In Sec. \ref{sec:res}, we present the results of confronting each of the metrics mentioned above with the available observational data. Finally, in Sec.  \ref{concl} we summarize our conclusion and discuss about future prospects.


\section{QPOs of the Lense-Thirring spacetime}\label{MetricLT}
According to the  relativistic precession model (RPM) \citep{1998ApJ...492L..59S}, the presence of three observed QPO frequencies is attributed to the Keplerian frequency, the precession frequency in the periastron, and the nodal precession frequency associated with the Lense-Thirring effect at a specific radius within the accretion disk. It has been demonstrated that these three frequencies, as predicted by the theory of General Relativity, can be described by a system of equations involving three unknowns: $M$ (the total mass), $J$ (the angular momentum) and $R$ (the radial location where the QPOs originate around the compact object) \citep{2014MNRAS.437.2554M} . If three QPOs are simultaneously observed, this system of equations can be analytically solved. Alternatively, when an independent measurement of the black hole mass is available (such as from dynamical spectrophotometric observations), together with two out of the three relevant QPOs for the RPM \citep{2014MNRAS.444.2065I,2014MNRAS.437.2554M}, numerical methods can be employed to find a solution \cite{2014GrCo...20..233B, 2015ARep...59..441B,2023PhRvD.108d4063B, 2022arXiv221210186B}.

If a spherical body rotates slowly enough with a uniform angular velocity along the axis of symmetry, the geometry around such an object can be well approximated by the Lense-Thirring solution.
\begin{eqnarray}
ds^2=-\left(1-\frac{2M}{r}\right)dt^{2}+\left(1-\frac{2M}{r}\right)^{-1}dr^{2}\nonumber\qquad\qquad\\
+r^2 \left(d\theta^{2}+\sin^2\theta d\phi^{2}\right)+\frac{4Ma}{r}\sin^{2}\theta d\phi dt,\
\label{eq:metricLT}
\end{eqnarray}
where $a=J/M$ is the angular momentum per unit mass (the Kerr parameter).
The components of the metric tensor are independent of the coordinates $t$ and $\phi$, which implies the existence of two constants of motion: the conserved specific energy per unit mass at infinity, $E$, and the conserved $z$ component of the specific angular momentum per unit mass at infinity, $L_z$. This circumstance allows us to write the $t$- and $\phi$-component of the 4-velocity of a test particle as \citep{2012JCAP...09..014B}
\be
u^t = \frac{E g_{\phi\phi} + L_z g_{t\phi}}{
g_{t\phi}^2 - g_{tt} g_{\phi\phi}} \, , \qquad
u^\phi = - \frac{E g_{t\phi} + L_z g_{tt}}{
g_{t\phi}^2 - g_{tt} g_{\phi\phi}} \, .
\label{utuphiKerr}
\ee
From the conservation of the rest-mass, $g_{\mu\nu}u^\mu u^\nu = -1$,
one can write
\be
g_{rr}\dot{r}^2 + g_{\theta\theta}\dot{\theta}^2
= V_{\rm eff}(r,\theta,E,L_z) \, ,
\ee
where $\dot{r} = u^r = dr/d\lambda$, $\dot{\theta} = u^\theta = d\theta/d\lambda$,
$\lambda$ is an affine parameter, and the effective potential $V_{\rm eff}$ is given by
\be
V_{\rm eff} = \frac{E^2 g_{\phi\phi} + 2 E L_z g_{t\phi} + L^2_z
g_{tt}}{g_{t\phi}^2 - g_{tt} g_{\phi\phi}} - 1  \, .
\ee
Circular orbits on the equatorial plane are located at the zeros and the turning points of the effective potential: $\dot{r} = \dot{\theta} = 0$, which implies $V_{\rm eff} = 0$, and $\ddot{r} = \ddot{\theta} = 0$, requiring respectively $\partial_r V_{\rm eff} = 0$ and $\partial_\theta V_{\rm eff} = 0$. From these conditions, one can obtain the Keplerian angular velocity $\Omega_{\phi}$, $E$ and $L_z$ of test particles:
\be
\Omega_{\phi} =
\frac{- \partial_r g_{t\phi}
\pm \sqrt{\left(\partial_r g_{t\phi}\right)^2
- \left(\partial_r g_{tt}\right) \left(\partial_r
g_{\phi\phi}\right)}}{\partial_r g_{\phi\phi}} \, ,
\ee
\be\label{E}
E = - \frac{g_{tt} + g_{t\phi}\Omega_{\phi}}{\sqrt{-g_{tt} - 2g_{t\phi}\Omega_{\phi} - g_{\phi\phi}\Omega_{\phi}^2}} \, ,
\ee
\be\label{L}
L_z = \frac{g_{t\phi} + g_{\phi\phi}\Omega_{\phi}}{\sqrt{-g_{tt} - 2g_{t\phi}\Omega_{\phi} - g_{\phi\phi}\Omega_{\phi}^2}} \, ,
\ee
where in $\Omega_{\phi}$ the sign ``$+$'' is for co-rotating (prograde) orbits and the sign ``$-$'' for counter-rotating (retrograde) ones. The Keplerian frequency is simply $f_{\phi} =\Omega_{\phi}/2\pi$.
The orbits are stable under small perturbations if $\partial_r^2 V_{\rm eff} \le 0$ and $\partial_\theta^2 V_{eff} \le 0$. In the case of more general orbits, we have that  \citep{2005Ap&SS.300..143K}
\be\label{eq-or}
\Omega^2_r =- \frac{1}{2 g_{rr} (u^t)^2}
\frac{\partial^2 V_{\rm eff}}{\partial r^2} \, , 
\ee
\be
\Omega^2_\theta = - \frac{1}{2 g_{\theta\theta} (u^t)^2}
\frac{\partial^2 V_{\rm eff}}{\partial \theta^2} \, .
\label{eq-ot}
\ee
The radial epicyclic frequency is thus $f_r = \Omega_r/2\pi$ and the vertical one is $f_\theta = \Omega_\theta/2\pi$.

The Keplerian angular velocity of a test particle in the Lense-Thirring spacetime takes the following form
\begin{equation}
\Omega_\phi^2\approx\frac{M}{r^3}\mp \frac{2\left(M/r\right)^{3/2}}{r^3}a+O(a^2)\ ,\label{omega-phi-LT}   
\end{equation}
and the corresponding frequency is $f_{\phi}=\Omega_\phi/2\pi$.
In addition, the radial and vertical angular frequencies are given by
\begin{equation}
 \Omega_r^2\approx\frac{M}{r^3}\left(1-\frac{6M}{r}\right)\pm \frac{6\left(M/r\right)^{3/2}(r+2M)}{r^4}a+O(a^2), \label{omega-r-LT}   
\end{equation}

\begin{equation}
\Omega_\theta^2\approx\frac{M}{r^3}\mp \frac{6\left(M/r\right)^{3/2}}{r^3}a+O(a^2)\  .\label{omega-theta-LT}
\end{equation}
The RPM identifies the lower QPO frequency $f_{L}$ with the periastron precession, namely $f_L=f_{\phi}-f_r$, and the upper QPO frequency $f_{U}$ with the Keplerian frequency, namely $f_{U}=f_{\phi}$.
Another physical quantity of great interest is the radius of the innermost stable circular orbit ($r_{ISCO}$). Using Eq. \eqref{E} or \eqref{L}, and imposing the condition $dE/dr=0$ or $dL/dr=0$ produces $r_{ISCO}$ which is given by
\be\label{iscoLT}
    r_{ISCO}=6M\left(1\mp\frac{2}{3}\sqrt{\frac{2}{3}}j\right)
\ee
where the $\mp$ signs correspond to co-rotating and counter-rotating particles and $j=a/M=J/M^2$ is the dimensionless angular momentum (spin parameter).

\section{QPOs of the Kerr spacetime}
\label{sez2b}
The Kerr spacetime is a vacuum solution to the field equations, describing a rotating uncharged black hole  \citep{1963PhRvL..11..237K}. The line element for the Kerr metric in the Boyer-Lindquist coordinates is given by
\begin{eqnarray}
ds^2=-\left(1-\frac{2Mr}{\Sigma}\right)dt^{2}+\frac{\Sigma}{\Delta}dr^{2}+\Sigma d\theta^{2}\nonumber\qquad\qquad\qquad\qquad\\
+\left(r^{2}+a^{2}+\frac{2Mra^{2}}{\Sigma}\sin^2\theta\right)\sin^2\theta d\phi^{2}-\frac{4Mra}{\Sigma}\sin^{2}\theta d\phi dt,
\label{eq:metricKerr}
\end{eqnarray}
where $\Sigma=r^{2}+a^{2}\cos^{2}\theta$ and $\Delta=r^{2}-2Mr+a^{2}$. The total gravitational mass of the source is given by $M$ and its angular momentum is $J=aM=jM^2$. For vanishing angular momentum the Kerr metric reduces to the Schwarzschild metric. 
The fundamental (epicyclic) frequencies of test particles in the Kerr spacetime are given by
\begin{equation}
   \Omega_\phi^2=\frac{M}{r^{3}\pm 2ar^{2}\sqrt{M/r}+a^{2}M}\ ,
\label{omegaKerr}
\end{equation}

\begin{equation}
 \Omega_r^{2}=\frac{M\left[r\left(r-6M)\right)\pm 8ar\sqrt{M/r}-3a^{2}\right]}{r^{2}\left(r^{3}\pm 2ar^{2}\sqrt{M/r}+a^{2}M\right)},   
\end{equation}
\begin{equation}
 \Omega_\theta^2=\frac{M\left(r^{2}\mp 4ar\sqrt{M/r}+3a^{2}\right)}{r^{2}\left(r^{3}\pm 2ar^{2}\sqrt{M/r}+a^{2}M\right)}.   
\end{equation}

The $r_{ISCO}$ is defined by \citep{1972ApJ...178..347B}
\begin{equation}
 \label{eq:riscoKerr}
\frac{r_{ISCO}^{\pm}}{M}=3+Z_2\pm \sqrt{(3-Z_1)(3+Z_1+2Z_2)},   
\end{equation}
with
\begin{subequations}
\begin{align}
Z_1&\equiv 1+\left(1-j^2\right)^{\frac{1}{3}}\left(\left(1+j\right)^{\frac{1}{3}}+\left(1-j\right)^{\frac{1}{3}}\right),\label{eq:Z1}\\ 
Z_2&\equiv \left(3 j^2+Z^2_1\right)^{\frac{1}{2}},\label{eq:Z2} 
\end{align}
\end{subequations}
where the $\pm$ signs correspond to counter-rotating  and co-rotating particles.


\section{QPOs of the  Zipoy-Voorhees spacetime}
\label{metric}

In spherical-like coordinates the Zipoy-Voorhees metric is represented by the line element

\begin{eqnarray}\label{line-element}
 ds^2=-f^\gamma
dt^2+f^{\gamma^2-\gamma}g^{1-\gamma^2}\left(\frac{dr^2}{f}+r^2d\theta^2\right)\nonumber\qquad\qquad\\
+f^{1-\gamma}r^2\sin^2\theta
d\phi^2\ ,   
\end{eqnarray}
where
\begin{equation}
    f(r)=1-\frac{2m}{r}\ , \qquad
g(r,\theta)=1-\frac{2m}{r}+\frac{m^2\sin^2\theta}{r^2}\ .\nonumber
\end{equation}

The real constant  $\gamma$ (sometimes denoted as $\delta$) is called the Zipoy-Voorhees parameter. If it is chosen as $\gamma=1+q$, where $q$ represents the  deformation of the source, i.e., the quadrupole parameter, the Zipoy-Voorhees metric is called  $q-$metric to emphasize the physical importance of the quadrupole parameter  \citep{quev12,Boshkayev2016}. 
The fundamental frequencies measured by observers at infinity are \citep{TMD}:
\begin{equation}
 \Omega_\phi^2=\frac{m\gamma(1-2m/r)^{2\gamma-1}}{r^2(r-m-m\gamma)}\ ,\label{omega-phi2}\\   
\end{equation}

\begin{eqnarray}
 \Omega_r^2=\frac{m\gamma(r-m)^{2\gamma^2-2} }{r^{(\gamma+1)^2}(r-2m)^{(\gamma-1)^2}(r-m-m\gamma)}\nonumber\qquad\qquad\quad\quad\\ 
\times\left[r^2-2(3\gamma+1)mr+2(\gamma +1)(2\gamma +1) m^2\right],\label{omega-r2}
\end{eqnarray}

\begin{equation}
\Omega_\theta^2=\frac{m\gamma(r-m)^{2\gamma^2-2} }{r^{\gamma^2+2\gamma}(r-2m)^{\gamma^2-2\gamma}(r-m-m\gamma)}\ .\label{omega-theta2} 
\end{equation}

For circular orbits, the ISCO radius can have two complementary values given by
\begin{equation}\label{isco}
  r_{ISCO}^{\pm}=m\left(1+3\gamma\pm\sqrt{5\gamma^2-1}\right)\,  .
\end{equation}

In order to estimate the effects due to the departure from spherical symmetry of the spacetime, one can study how small deviations from the Schwarzschild metric affect the epicyclic frequencies. By considering $\gamma=1+q$ with $|q|\ll1$, the epicyclic frequencies take the following form

\begin{align}
\Omega_\phi&\approx\Omega_{\phi,0}\left[1+\left\{\frac{r-m}{2(r-2m)}+\log\left(1-\frac{2m}{r}\right)\right\}q+ O(q^2) \right],\label{omega-phi-small}
\end{align}
\begin{align}
\Omega_r&\approx\Omega_{r,0}\left[1+\left\{\frac{r^2-13mr+20 m^2}{2(r-2m)(r-6m)}+\log\left(1-\frac{m}{r}\right)^2\right\} q \right. \nonumber\\
  &\left. + O(q^2) \right],\label{omega-r-small}
\end{align}
\begin{align}
\Omega_\theta&\approx\Omega_{\theta,0}\left[1+ \left\{\frac{r-m}{2(r-2m)}+\log\left(1-\frac{m}{r}\right)^2\right\}q+ O(q^2) \right],\label{omega-theta-small}
\end{align}
where the subscript ``0'' corresponds to the frequencies in the Schwarzschild spacetime
\begin{equation}
\Omega_{\theta,0}=\Omega_{\phi,0}=\sqrt{\frac{m}{r^3}}\ , 
\end{equation}
\begin{equation}
  \Omega_{r,0}=\sqrt{\frac{m}{r^3}\left(1-\frac{6m}{r}\right)}\ .
\end{equation}

Thus, one can see from Eqs.~(\ref{omega-r-small}),~(\ref{omega-theta-small}), and (\ref{omega-phi-small}) that values of $q>0$ ($q<0$) represent oblate (prolate) deviations from spherical symmetry, which alter the values of all components of the epicyclic frequency of test particles.

The ISCO radius is 
\begin{equation}
r_{ISCO,\pm} 
=\begin{cases}
   m(6+11q/2)\ , &\text{for ``+''}\\
   m(2+q/2)\ , &\text{for ``-''}
 \end{cases}.
\end{equation}
We will focus on the ``+'' case as the case with ``-'' sign turns out to be non-physical. 

Moreover, the Geroch-Hansen multipole moments \citep{ger,ger2,hans} of  the $\gamma$ -metric are given by
\begin{eqnarray}\label{gerhanmomgamma}
 M_{2k+1}=0, \qquad k=0, 1, 2, ... ,\nonumber\qquad\qquad\qquad\qquad\qquad\qquad\\  
 M_0=m\gamma, \qquad M_2=-\frac{1}{3}m^3(-1+\gamma)\gamma(1+\gamma).\qquad\qquad\quad  
\end{eqnarray}
In case of the $q$-metric, they are
\begin{equation}\label{eqmm15}
 M_0=m(1+q), \quad M_2=-\frac{1}{3}m^3q(1+q)(2+q).   
\end{equation}
For vanishing $q$ or equivalently $\gamma=1$, one recovers multipole moments for the Schwarzschild metric.


\begin{table*}
\centering
\setlength{\tabcolsep}{0.65em}
\renewcommand{\arraystretch}{1.3}
\begin{tabular}{lccccrrrrr}
\hline\hline
Source                                  &
Metric                                  &
$M$ $({\rm M}_\odot)$                   &
$q$                                     &
$j$                                     &
$\mathcal L_{\rm max}$                  &
AIC                                     & 
BIC                                     &
$\Delta$AIC                             &
$\Delta$BIC                             \\
\hline
Cir X1                                  &
S                                       &
$2.224^{+0.029}_{-0.029}$               &
--                                      &
--                                      &
$-125.84$ & $254$ & $254$ & $117$ & $117$  \\
                                        &
ZV                                      &
$2.379^{+0.059}_{-0.056}$               &
$-0.97^{+0.15}_{-0.03}$                 &
--                                      &
$-114.17$ & $233$ & $233$ & $96$ & $96$  \\
                                        &
K                                       &
$5.12^{+0.15}_{-0.24}$                  &
--                                      &
$0.995^{+0.005}_{-0.028}$               &
$-77.39$ & $159$ & $159$ & $22$ & $22$  \\
                                        &
LT                                      &
$3.870^{+0.077}_{-0.097}$               &
--                                      &
$0.573^{+0.014}_{-0.015}$               &
$-66.32$ & $137$ & $137$ & $0$ & $0$   \\
\hline
GX 5-1                                  &
S                                       &
$2.161^{+0.010}_{-0.010}$               &
--                                      &
--                                      &
$-200.33$ & $402$ & $404$ & $115$ & $115$  \\
                                        &
ZV                                      &
$1.978^{+0.019}_{-0.016}$               &
$0.979^{+0.021}_{-0.099}$               &
--                                      &
$-168.96$ & $342$ & $344$ & $55$ & $55$  \\
                                        &
K                                       &
$1.286^{+0.027}_{-0.014}$               &
--                                      &
$-0.986^{+0.048}_{-0.014}$              &
$-141.66$ & $287$ & $289$ & $0$ & $0$   \\
                                        &
LT                                      &
$1.86^{+0.10}_{-0.12}$                  &
--                                      &
$-0.26^{+0.10}_{-0.014}$                &
$-192.89$ & $390$ & $392$ & $103$ & $103$  \\
\hline
GX 17+2                                 &
S                                       &
$2.0768^{+0.0002}_{-0.0003}$            &
--                                      &
--                                      &
$-1819.02$ & $3641$ & $3642$ & $3530$ & $3530$  \\
                                        &
ZV                                      &
$1.9569^{+0.0038}_{-0.0043}$            &
$0.996^{+0.004}_{-0.017}$               &
--                                      &
$-1511.25$ & $3027$ & $3028$ & $2916$ & $2916$  \\
                                        &
K                                       &
$8.60^{+0.14}_{-0.16}$                  &
--                                      &
$0.9961^{+0.0039}_{-0.0048}$            &
$-53.26$ & $111$ & $112$ & $0$ & $0$   \\
                                        &
LT                                      &
$1.4451^{+0.0044}_{-0.0044}$            &
--                                      &
$-0.9982^{+0.0076}_{-0.0018}$           &
$-661.44$ & $1327$ & $1328$ & $1216$ & $1216$  \\
\hline
GX 340+0                                &
S                                       &
$2.1023^{+0.0033}_{-0.0033}$            &
--                                      &
--                                      &
$-130.86$ & $264$ & $265$ & $10$ & $10$  \\
                                        &
ZV                                      &
$1.951^{+0.041}_{-0.027}$               &
$0.81^{+0.17}_{-0.25}$                  &
--                                      &
$-125.19$ & $255$ & $256$ & $1$ & $1$  \\
                                        &
K                                       &
$1.56^{+0.16}_{-0.12}$                  &
--                                      &
$-0.52^{+0.18}_{-0.17}$                 &
$-124.79$ & $254$ & $255$ & $0$ & $0$   \\
                                        &
LT                                      &
$1.82^{+0.16}_{-0.13}$                  &
--                                      &
$-0.25^{+0.15}_{-0.15}$                 &
$-129.06$ & $262$ & $263$ & $9$ & $9$  \\
\hline
Sco X1                                  &
S                                       &
$1.96490^{+0.00054}_{-0.00054}$         &
--                                      &
--                                      &
$-3887.17$ & $7775$ & $7778$ & $7507$ & $7507$  \\
                                        &
ZV                                      &
$2.02436^{+0.00038}_{-0.00042}$         &
$-0.1969^{+0.0025}_{-0.0022}$           &
--                                      &
$-2761.42$ & $5528$ & $5531$ & $5260$ & $5260$  \\
                                        &
K                                       &
$6.352^{+0.062}_{-0.068}$               &
--                                      &
$0.9274^{+0.0033}_{-0.0037}$            &
$-131.68$ & $268$ & $271$ & $0$ & $0$   \\
                                        &
LT                                      &
$1.2950^{+0.0013}_{-0.0010}$            &
--                                      &
$-0.9994^{+0.00279}_{-0.00061}$         &
$-1509.67$ & $3024$ & $3027$ & $2756$ & $2756$  \\
\hline
4U1608–52                               &
S                                       &
$1.9596^{+0.0038}_{-0.0038}$            &
--                                      &
--                                      &
$-235.83$ & $473$ & $474$ & $347$ & $347$  \\
                                        &
ZV                                      &
$2.0132^{+0.0022}_{-0.0030}$            &
$-0.165^{+0.014}_{-0.018}$              &
--                                      &
$-192.25$ & $389$ & $390$ & $263$ & $263$  \\
                                        &
K                                       &
$5.94^{+0.30}_{-0.26}$                  &
--                                      &
$0.906^{+0.019}_{-0.018}$               &
$-60.96$ & $126$ & $127$ & $0$ & $0$   \\
                                        &
LT                                      &
$1.302^{+0.021}_{-0.010}$               &
--                                      &
$-0.991^{+0.048}_{-0.010}$              &
$-131.77$ & $268$ & $269$ & $142$ & $142$  \\
\hline
4U1728–34                               &
S                                       &
$1.7345^{+0.0028}_{-0.0029}$            &
--                                      &
--                                      &
$-212.61$ & $427$ & $427$ & $351$ & $351$  \\
                                        &
ZV                                      &
$1.8444^{+0.0028}_{-0.0033}$            &
$-0.337^{+0.012}_{-0.011}$              &
--                                      &
$-111.79$ & $228$ & $228$ & $152$ & $152$  \\
                                        &
K                                       &
$6.57^{+0.30}_{-0.32}$                  &
--                                      &
$0.981^{+0.011}_{-0.014}$               &
$-36.07$ & $76$ & $76$ & $0$ & $0$   \\
                                        &
LT                                      &
$1.141^{+0.018}_{-0.010}$               &
--                                      &
$-0.989^{+0.048}_{-0.011}$              &
$-133.95$ & $272$ & $272$ & $196$ & $196$  \\
\hline
4U0614+091                              &
S                                       &
$1.9042^{+0.0014}_{-0.0014}$            &
--                                      &
--                                      &
$-842.97$ & $1687$ & $1689$ & $1372$ & $1372$  \\
                                        &
ZV                                      &
$1.726^{+0.010}_{-0.003}$               &
$0.986^{+0.014}_{-0.071}$               &
--                                      &
$-862.04$ & $1728$ & $1730$ & $1413$ & $1413$  \\
                                        &
K                                       &
$7.708^{+0.043}_{-0.139}$               &
--                                      &
$0.9989^{+0.0011}_{-0.0042}$            &
$-155.32$ & $315$ & $317$ & $0$ & $0$   \\
                                        &
LT                                      &
$1.2498^{+0.0061}_{-0.0026}$            &
--                                      &
$-0.9970^{+0.0154}_{-0.0030}$           &
$-502.38$ & $1009$ & $1011$ & $694$ & $694$  \\
\hline
\end{tabular}
\caption{Best-fit parameters got from MCMC simulations for Schwarzschild (S), Zipoy-Voorhees (ZV), Kerr (K) and Lense-Thirring (LT) metrics. $\Delta$AIC and $\Delta$BIC are computed with respect to the reference model.}
\label{tab:results_MCMC}
\end{table*}


\begin{figure*}
{\hfill
\includegraphics[width=0.41\hsize,clip]{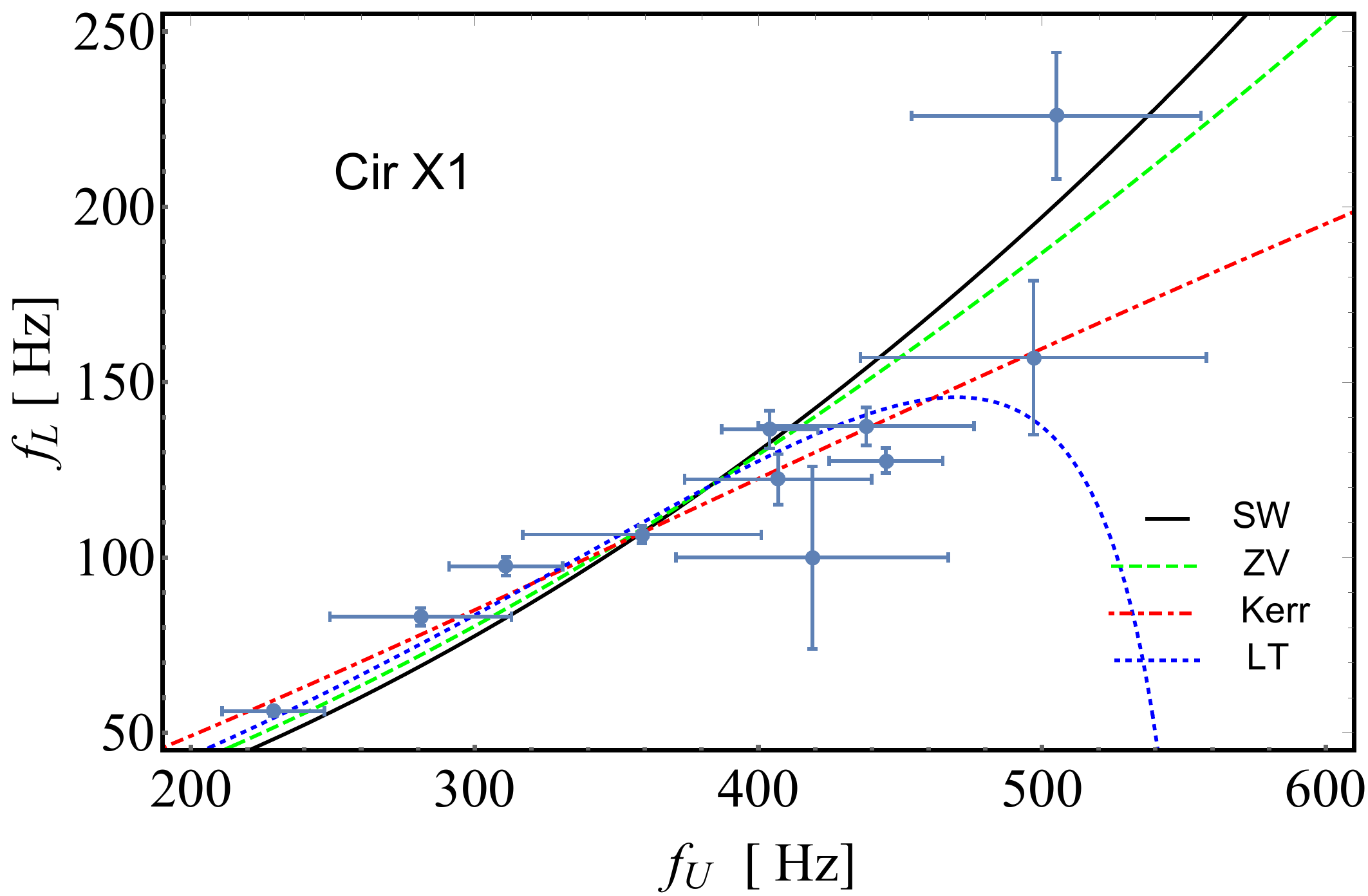}\hfill
\includegraphics[width=0.41\hsize,clip]{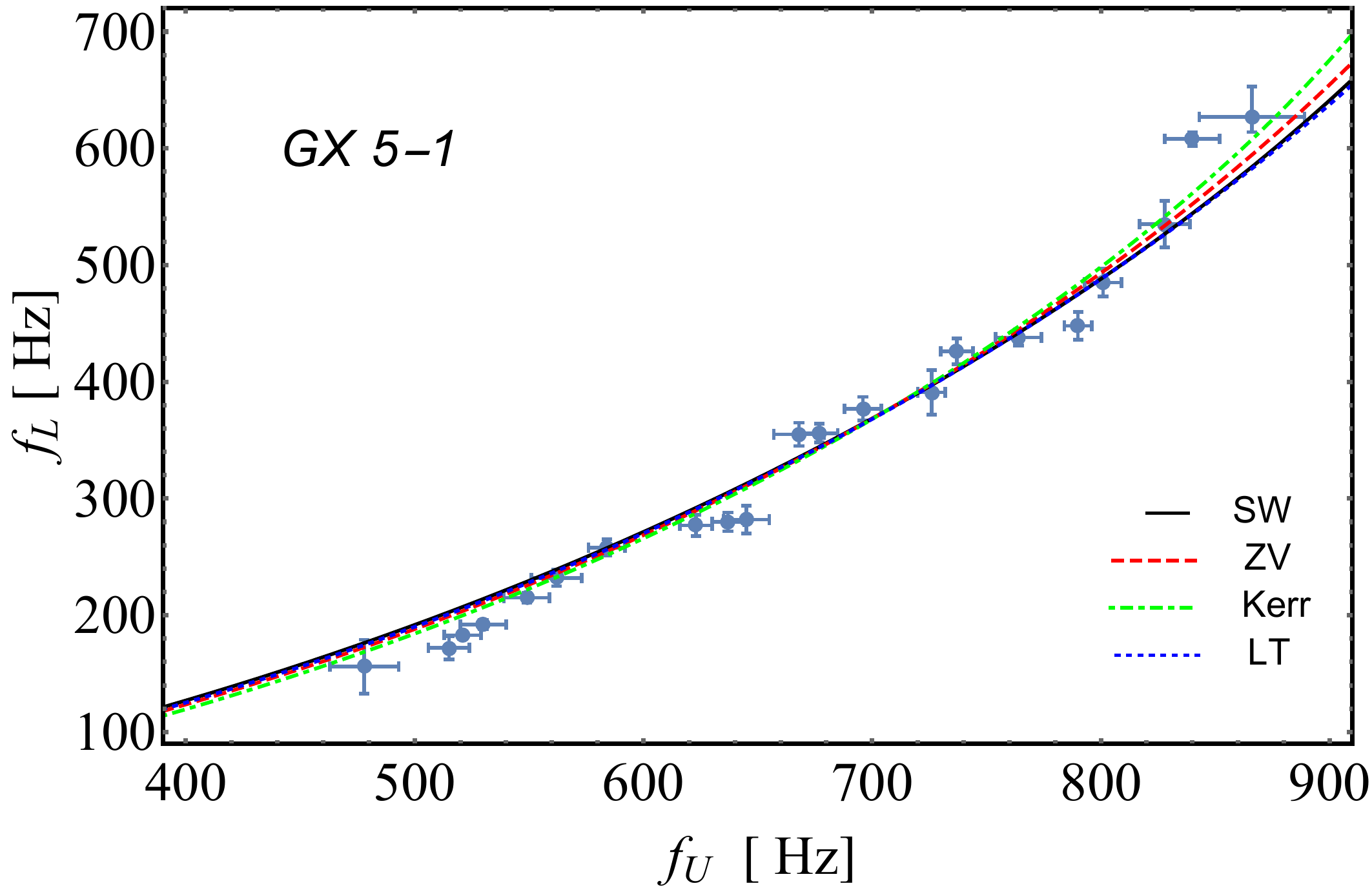}\hfill}

{\hfill
\includegraphics[width=0.41\hsize,clip]{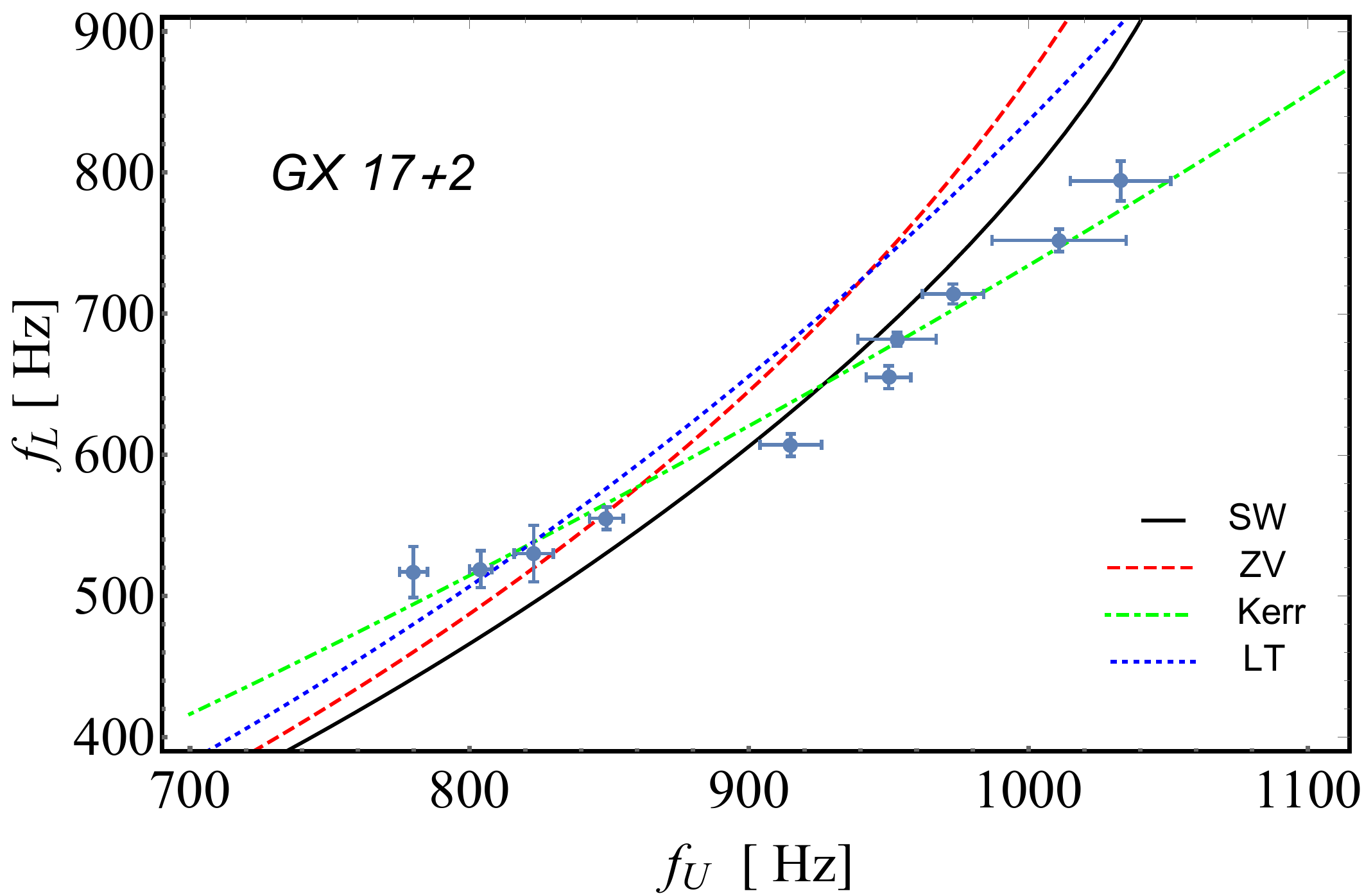}\hfill
\includegraphics[width=0.41\hsize,clip]{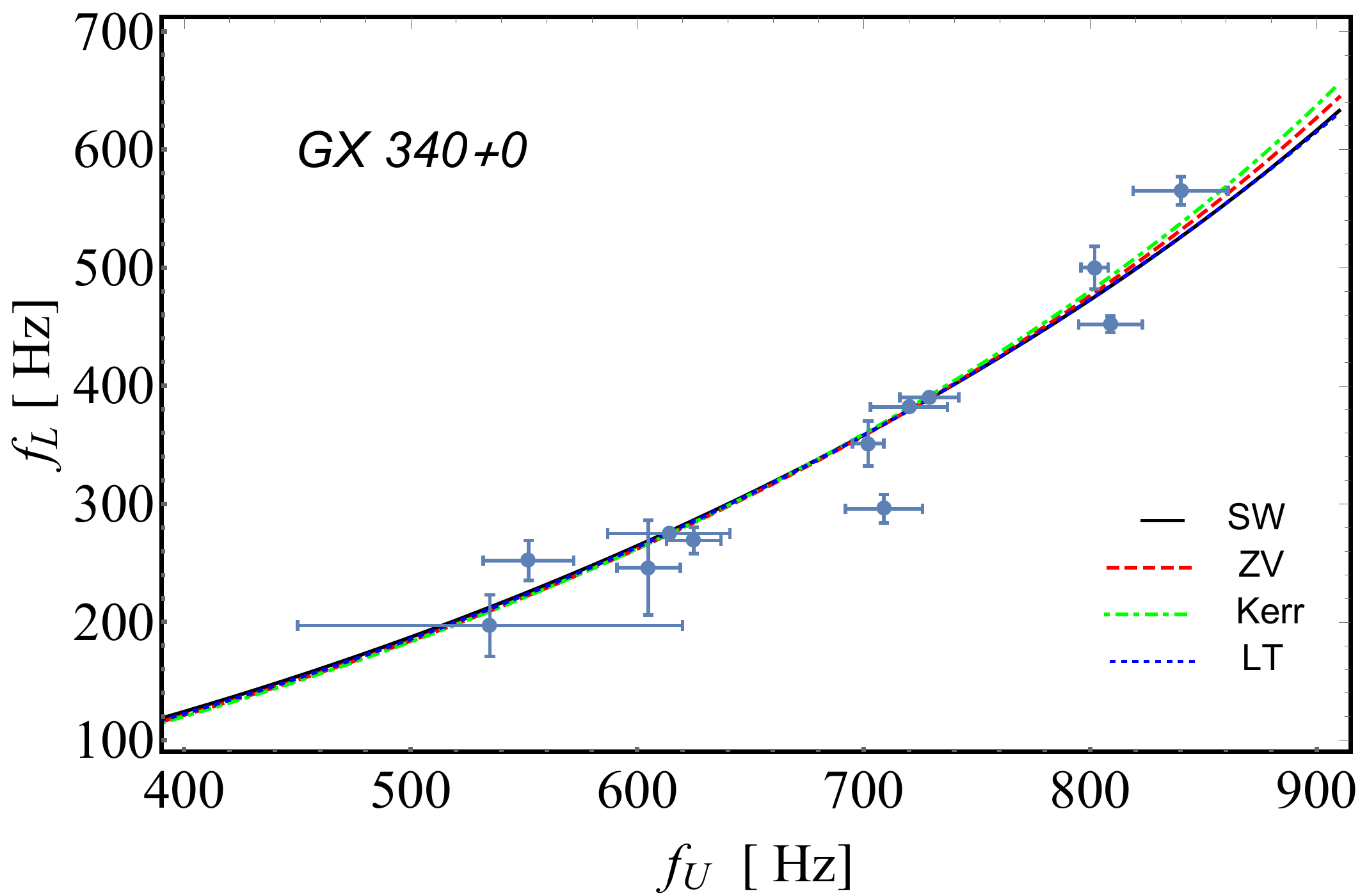}\hfill}

{\hfill
\includegraphics[width=0.41\hsize,clip]{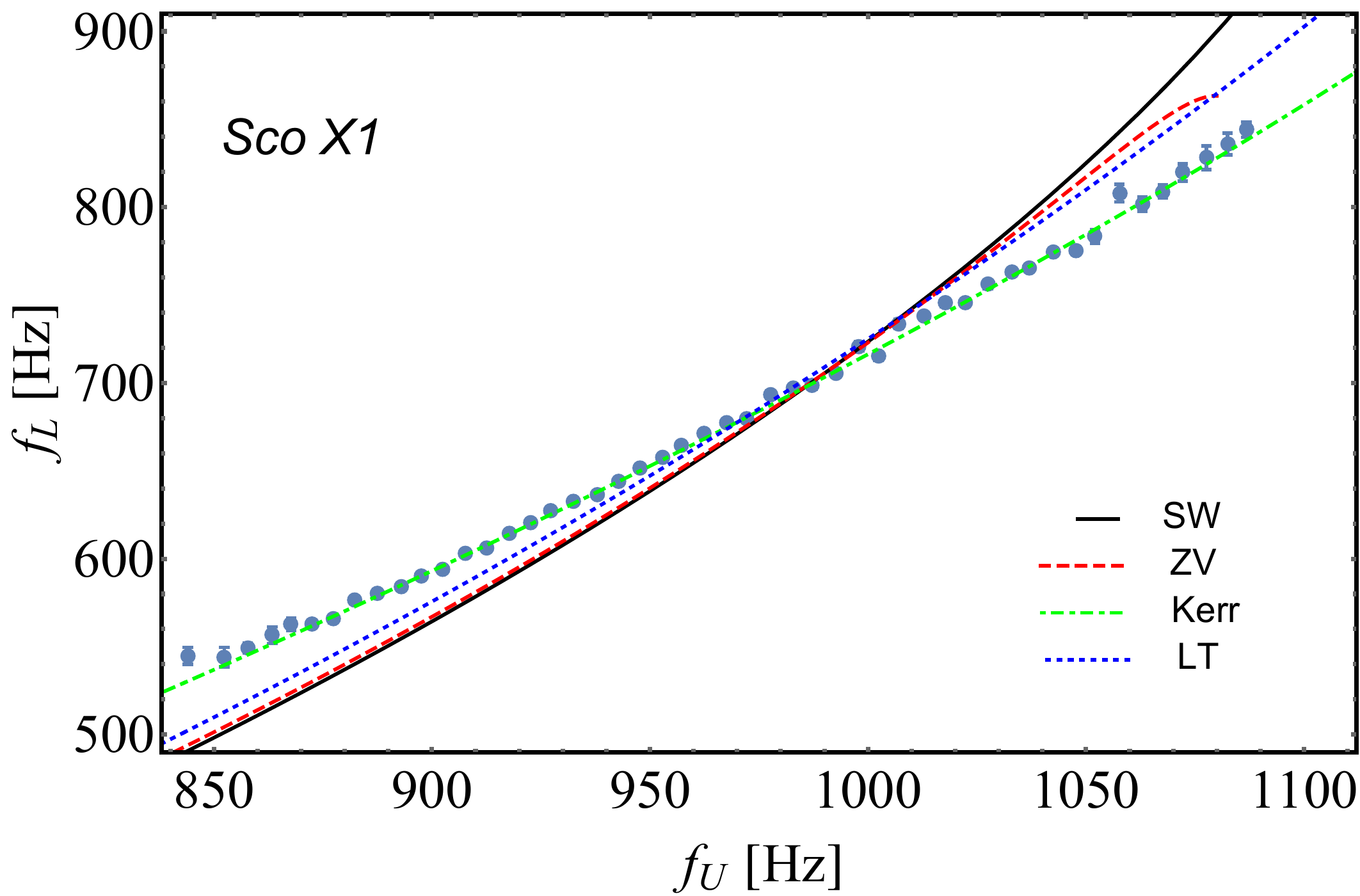}\hfill
\includegraphics[width=0.41\hsize,clip]{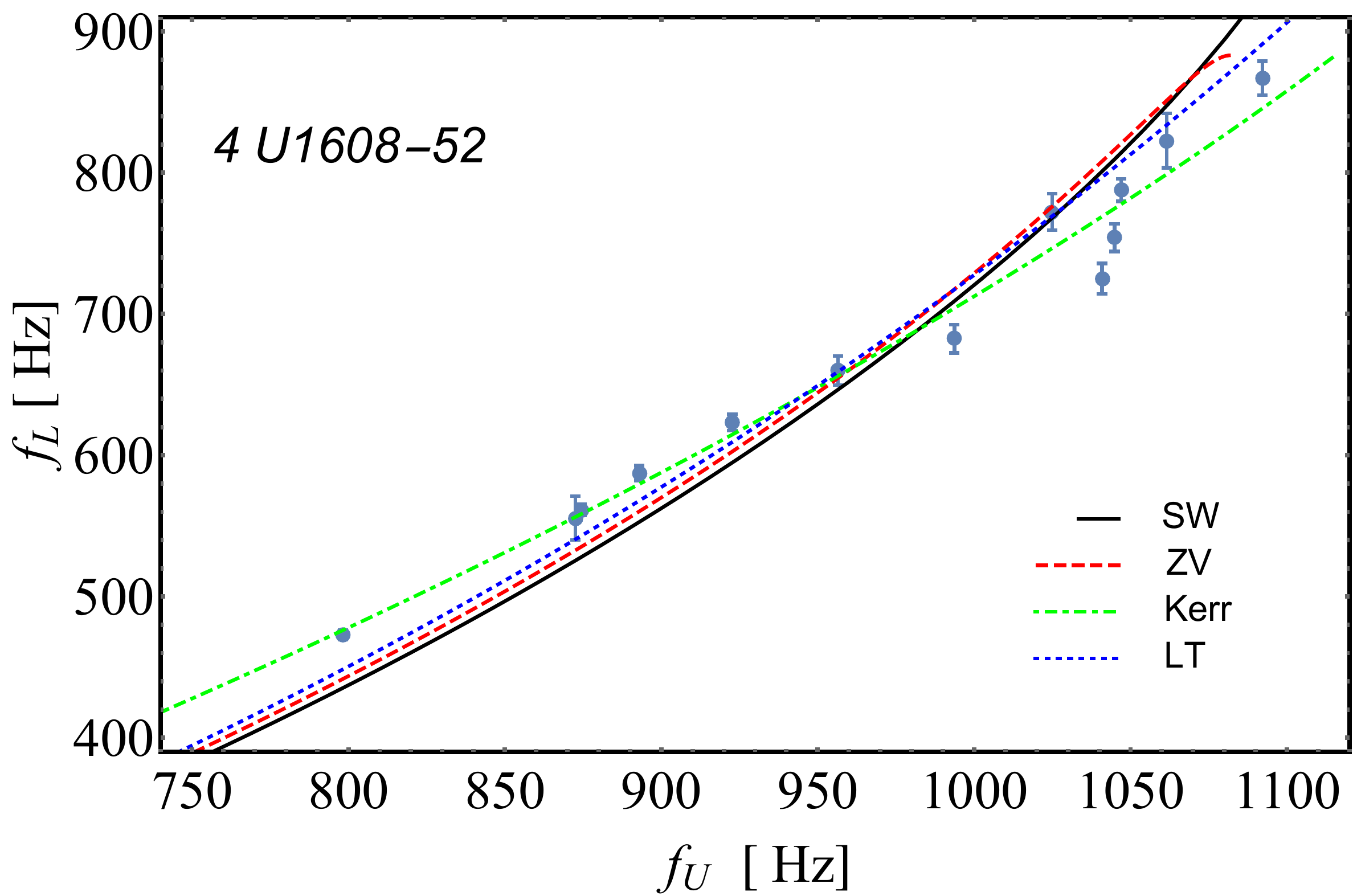}\hfill}

{\hfill
\includegraphics[width=0.41\hsize,clip]{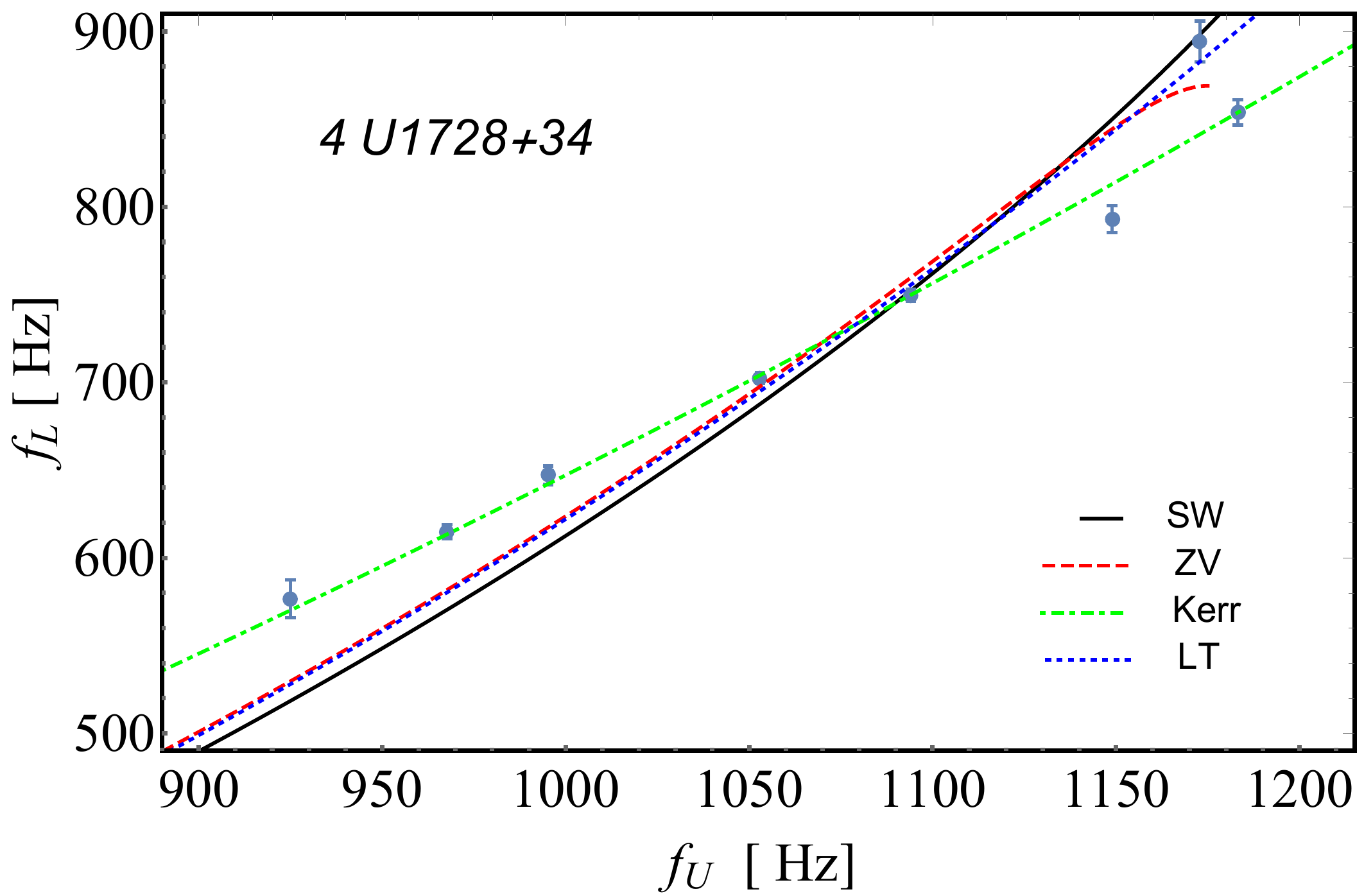}\hfill
\includegraphics[width=0.41\hsize,clip]{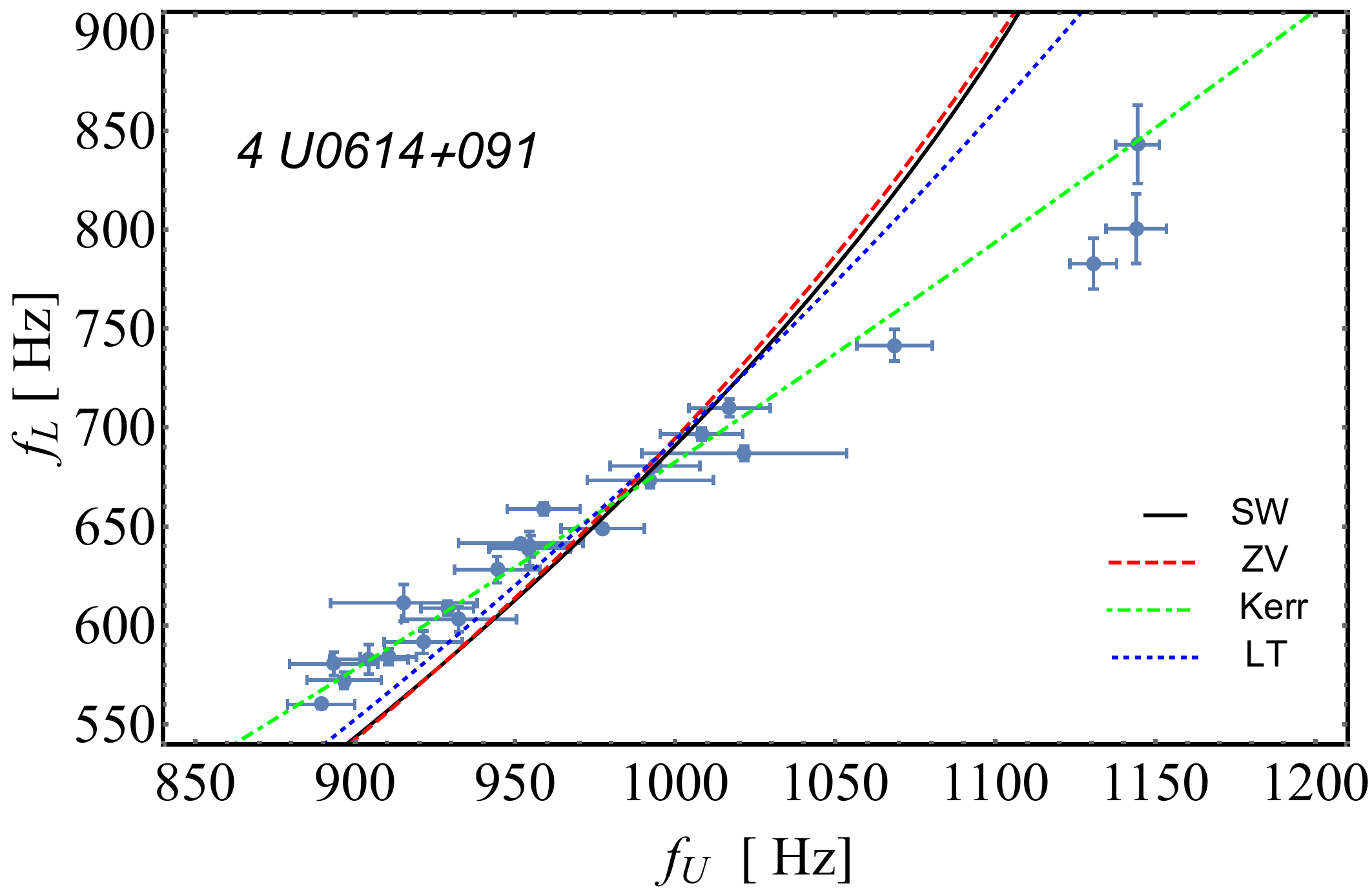}\hfill}
\caption{Plots of $f_{\rm L}$ vs $f_{\rm U}$ frequencies of the QPO data sets considered in this work (dark cyan data with error bars).}
\label{fig:freq}
\end{figure*}


\section{Numerical results and discussion}
\label{sec:res}

In this section, we discuss some physical consequences based on the information summarized in Tab.~\ref{tab:results_MCMC}. There, for each source, we establish the best-fit model out of the four metrics considered in this work.

To compare each theoretical model with the numerical data, we adjusted the free \texttt{Wolfram Mathematica} code from \citet{2019PhRvD..99d3516A} to carry out our MCMC analysis, which is based on the Metropolis-Hastings algorithm. We then identified the best-fit parameters by locating the peak of the log-likelihood, which is expressed as
\begin{equation}
\label{loglike}
    \ln L = -\sum_{k=1}^{N}\left\{\dfrac{\left[f_{\rm L}^k-f_{\rm L}(p,f_{\rm U}^k)\right]^2}{2(\sigma f_{\rm L}^k)^2} + \ln(\sqrt{2\pi}\sigma f_{\rm L}^k)\right\}\,,
\end{equation}
where $p$ labels the model parameters and $N$ the data for each source, sampled as lower frequencies $f_{\rm L}^k$, attached errors $\sigma f_{\rm L}^k$, and upper frequencies $f_{\rm U}^k$.
Not all the metrics considered in this work have analytic expressions for $f_{\rm L} =f_{\rm L}(p,f_{\rm U})$.

To compare each fit, based on different metrics, we single out two main selection criteria, {\emph i.e.}, the well-consolidated \emph{Aikake Information Criterion} (AIC) and \emph{Bayesian Information Criterion} (BIC) \citep{2007MNRAS.377L..74L}, respectively
\begin{subequations}
\begin{align}
{\rm AIC}&=-2\ln L+2p\,,\\
{\rm BIC}&=-2\ln L+p\ln N\,.
\end{align}
\end{subequations}
For each fit, the maximum value of the log-likelihood $\ln L$ is reported in Table \ref{tab:results_MCMC}. The fiducial (most suitable) model is the one with the lowest values of the AIC and BIC tests, namely ${\rm AIC}_0$ and ${\rm BIC}_0$. The other models are compared to the fiducial one by the differences $\Delta{\rm AIC}={\rm AIC}-{\rm AIC}_0$ and $\Delta{\rm BIC}={\rm BIC}-{\rm BIC}_0$.

When contrasting models, the proof against the suggested model or, in other words, in support of the reference model can be simply summarized as follows:
\begin{itemize}
    \item[-] $\Delta{\rm AIC}$ or $\Delta{\rm BIC}\in[0,\,3]$, weak evidence;
    \item[-] $\Delta{\rm AIC}$ or $\Delta{\rm BIC}\in (3,\,6]$, mild evidence;
    \item[-] $\Delta{\rm AIC}$ or $\Delta{\rm BIC}>6$, strong evidence.
\end{itemize}

For each QPO source, we analyze the data from statistical, observational, and physical perspectives. In what follows, we stress that although depending on the equation of state of neutron stars, theoretically one can have masses up to 3.2 $M_\odot$ \citep{2002BASI...30..523S}, from observations we now get up to 2.14 $M_\odot$ \citep{2020NatAs...4...72C}. We will \emph{disentangle} this concept in our theoretical interpretations below.

\begin{itemize}
\item[-] {\bf Cir~X-1} \citep{2006ApJ...653.1435B}. Cir X-1 is one of the most puzzling X-ray binaries known. From  Tab. \ref{tab:results_MCMC}, we see that statistically the LT metric is strongly preferred over the other ones.  
For the LT metric we inferred the mass $M=3.87 M_\odot$ and dimensionless angular momentum $0.573$. The value of the dimensionless angular momentum is in agreement with the theoretical constraints reported in \citet{2011ApJ...728...12L} and \citet{2016RAA....16...60Q}, and align with is the one reported in \citet{2010ApJ...714..748T}, where the ranges of mass and angular momentum were restricted to $M \in (1, 4) M_\odot$ and $j \in (0, 0.5)$. In addition, the mass $M=3.87 M_\odot$ is beyond observational and theoretical constraints.
\item[-] {\bf GX~5-1} \citep{1998ApJ...504L..35W, 2002MNRAS.333..665J}. Examining Table \ref{tab:results_MCMC} for the source GX 5-1, it is evident that statistically the Kerr metric yields the best fit. However, physically $j=-0.986$ is too large for NSs according to \citet{2011ApJ...728...12L} and the negative sign implies that the accretion disk rotates in opposite direction relative to the central object, which is, in principle, quite possible. Even though the mass $M=1.286$ aligns with the most of neutron star models and the value of the spin parameter $j=-0.986$ can be justified by the absence of the crust of NSs \citep{2016RAA....16...60Q}. However, it is known that the geometry around a NS is different from the Kerr spacetime \citep{1999ApJ...512..282L}. 
\item[-] {\bf GX~17+2} \citep{2002ApJ...568..878H}. In this case, statistically the Kerr metric provides the best fit, but, physically only the Schwarzschild metric is in agreement with NS models.
\item[-] {\bf GX~340+0} \citep{2000ApJ...537..374J}. Here, the Kerr and ZV metrics provide best fits, and the LT metric is weakly favored with respect to Kerr. However, the quadrupole parameter in the ZV metric does not satisfy the assumption $q\ll1$. Therefore, for this source the Kerr metric statistically gives best fit, though physically the LT and Schwarzschild metrics align with the NS physics.
\item[-] {\bf Sco~X1} \citep{2000MNRAS.318..938M} and {\bf 4U1608-52} \citep{1998ApJ...505L..23M}. For these sources the Kerr metric statistically produces best fits. The ZV metric gives contradicting sign for the quadrupole parameter ($q<0$), since one expects to have an oblate star relative to the axis of symmetry. Moreover, though the LT solution provides ``good'' masses for NSs, which are close to the canonical NS mass $M=1.4M_\odot$, the value of the spin parameter is quite large for NSs according to \citet{2011ApJ...728...12L}, but is fine with respect to \citet{2016RAA....16...60Q} if a NS does not have a crust and rotate fast. 
\item[-] {\bf 4U1728-34} \citep{1999ApJ...517L..51M}. Here the Kerr metric gives the best fit, but a more realistic physical description is provided by the Schwarzschild metric.
\item[-] {\bf 4U0614+091} \citep{1997ApJ...486L..47F}. Also for this source, the Kerr metric provides the best fit to the data. Nonetheless, physically only the Schwarzschild metric is viable solution, even though statistically not preferable.
\end{itemize}

The results are summarized in Table \ref{tab:results_MCMC}. The contour plots of the best-fitting model parameters are shown in Figs. \ref{fig:contoursZV}-\ref{fig:contoursLT} (see Appendix \ref{appA}). For each source, the dependence of the lower frequencies $f_{L}$ on the upper frequencies $f_{U}$ of the QPO data, for the four metrics considered in this work, are shown in Fig. \ref{fig:freq}. 


\section{Conclusion}
\label{concl}

In this paper, we conducted a numerical analysis of the QPO data for eight sources from LMXRs. All fit results underwent both statistical and theoretical analyses and were compared with observational constraints.

From a statistical standpoint, the Kerr metric provided the best fits for seven out of eight sources. In most cases, specifically for six sources, the Kerr metric yielded large masses and spin parameters within the range of $M \sim 5.12-8.6 M_\odot$ and $j \sim 1$. However, for only two sources, notably GX 5-1 and GX 340+0, the Kerr metric produced realistic masses of approximately $1.29 M_\odot$ and $1.56 M_\odot$, respectively. Additionally, only for GX 340+0 did the Kerr metric provide a reasonable value for the spin parameter, with $j=-0.52$, indicating counter-rotating orbits.

It is essential to highlight that the Kerr metric was intentionally chosen to assess the nature of compact companions in the LMXRs, which are anticipated to be NSs. It is well-established that the Kerr metric is not suitable for describing the geometry around NSs. Consequently, we were prepared {\it a priori} for the expectation that the Kerr metric might yield non-physical results. The exceptions were limited to the two aforementioned sources, and only one of them satisfied all physical and observational constraints.

Based on the fits, our expectation was that the masses of the NSs would be around the maximum observed mass of $2.35 M_\odot$ \citep{2022ApJ...934L..17R} and/or around the maximum theoretical mass of approximately $\sim 3M_\odot$ \citep{2002BASI...30..523S,2014PhRvC..90d4305C,2014NuPhA.921...33B,2020Univ....6..119B}. Adopting $\sim 3M_\odot$ as a reference, the Kerr metric fails to align with these observational constraints on mass for six sources. However, the situation with the spin parameter corresponding to the masses was in accordance with \citet{2011ApJ...728...12L,2018mgm..conf.3433B, 2016RAA....16...60Q} for only two sources. We conclude that the Kerr metric is indeed unsuitable for describing six sources based on our analyses.

It is well-known that the geometry around rotating NSs can be approximated by the Kerr metric only in the regime of slow rotation, equivalent to the LT metric. Therefore, the LT metric was also considered in our analysis. Ultimately, the LT metric provided the best fit for only one source, Cir X1, although the mass exceeds both observational and theoretical constraints. Additionally, statistically, the LT metric ranked second after the Kerr metric for the following sources: GX 17+2, Sco X1, 4U1608-52, 4U0614+091. For these sources, the masses align with both observation and theory, although the values of the spin parameter exceed theoretical constraints \citep{2011ApJ...728...12L,2016RAA....16...60Q, 2018mgm..conf.3433B} for the given masses.

The ZV metric follows the Kerr metric for the sources GX 5-1, GX 340+0, 4U1728-34. While the cases involving mass were ideal, the cases with the quadrupole parameter yielded contradicting results, violating the crucial conditions $q\ll1$ and $q>0$. It should be noted that we did not use the exact ZV metric. In order to obtain an analytic fitting function, the ZV metric was expanded into a Taylor series for small $q$. In future analyses, we aim to employ the exact expression for the ZV metric.

The Schwarzschild metric consistently provided the worst fit. This may be attributed to the fact that it contains only one free parameter relative to other metrics. Nevertheless, from physical and observational perspectives, the Schwarzschild metric yields viable and reliable masses.

Unfortunately, direct measurements of the masses of compact components in LMXBs are still unavailable. This lack poses significant challenges in testing theoretical models of NSs. Observational constraints, which may deviate from theoretical predictions, are primarily derived from a limited number of sources \citep{2011PrPNP..66..674T}. Consequently, it is not always feasible to directly compare our findings with observational data.

There could be several reasons for obtaining contradicting statistical results with physical (theoretical) analyses:
\begin{itemize}
\item[-] The Kerr metric describes limiting class of objects, namely BHs, and it is not suitable for the description of the geometry around NSs.
\item[-] The ZV metric describes static deformed objects, but in general NSs rotate. Only in the limit of slow rotation, the geometry can be approximated by small deformation due to mimicking effects. As we have noticed from our fits, the rotation was not slow, and the deformation was not small.
\item[-] As mentioned above, NSs not only rotate, but also deviate from a spherical symmetry, so the metric must include at least three parameters for the source: the total mass, angular momentum and quadrupole moment.
Exact solutions with these parameters exist in the literature. We expect to continue exploring such solutions in future works.
\item[-]  There could be effects of the magnetic and induced electric fields of NSs on the motion of both neutral and charged particles in the accretion disks.
\item[-]  The RPM model is the simplest model for QPOs. Since the mechanism for generating QPOs is not well understood, there could be extra effects which should be taken into consideration.
\end{itemize}

Consequently, we conclude that more thorough and extensive analyses are required for these sources, involving other solutions to the field equations and QPO models.

\section*{Acknowledgments} TK acknowledges the Grant No. AP19174979, YeK acknowledges Grant No. AP19575366, while KB and MM acknowledge Grant No. BR21881941 from the Science Committee of the Ministry of Science and Higher Education of the Republic of Kazakhstan. KB expresses his gratitude to Professor Mariano M{\'e}ndez for providing data points for QPOs.

\section*{Data availability}
The data points for quasiperiodic oscillations have been retrieved from publicly available papers:
\cite{2006ApJ...653.1435B},
\cite{1998ApJ...504L..35W}, \cite{2002MNRAS.333..665J},
\cite{2002ApJ...568..878H},
\cite{2000ApJ...537..374J},
\cite{2000MNRAS.318..938M},
\cite{1998ApJ...505L..23M},
\cite{1999ApJ...517L..51M},
\cite{1997ApJ...486L..47F}.
In addition, for some sources the data points were kindly provided by Professor Mariano M{\'e}ndez via private communication.

\appendix

\section{Contour plots}\label{appA}
\begin{figure*}
{\hfill
\includegraphics[width=0.31\hsize,clip]{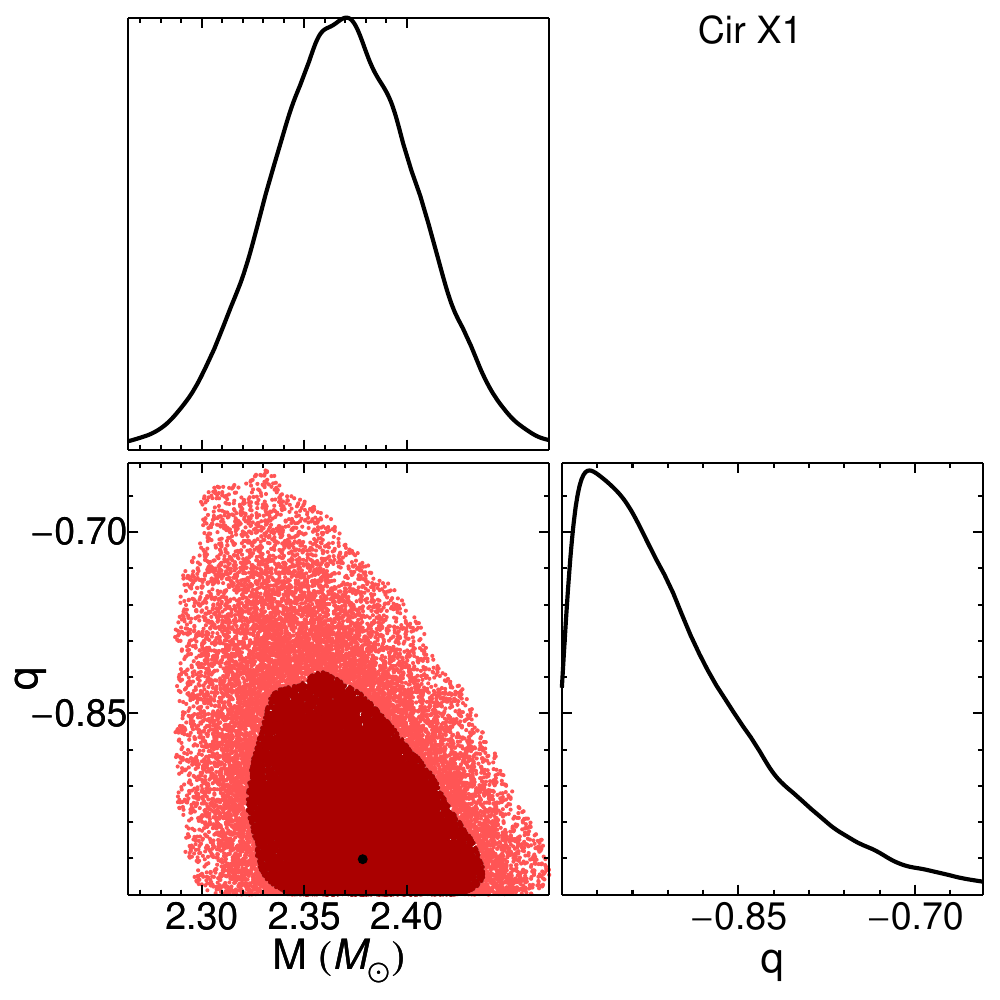}\hfill
\includegraphics[width=0.31\hsize,clip]{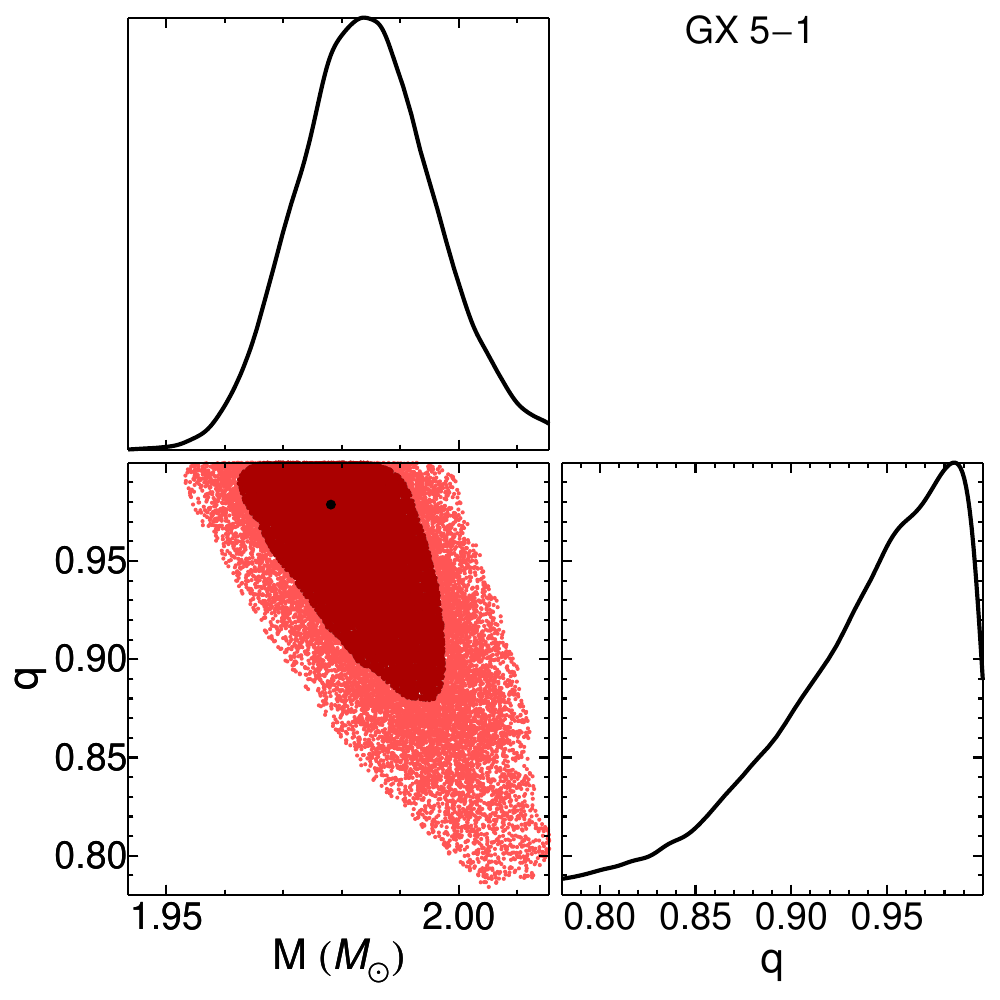}
\hfill}

{\hfill
\includegraphics[width=0.31\hsize,clip]{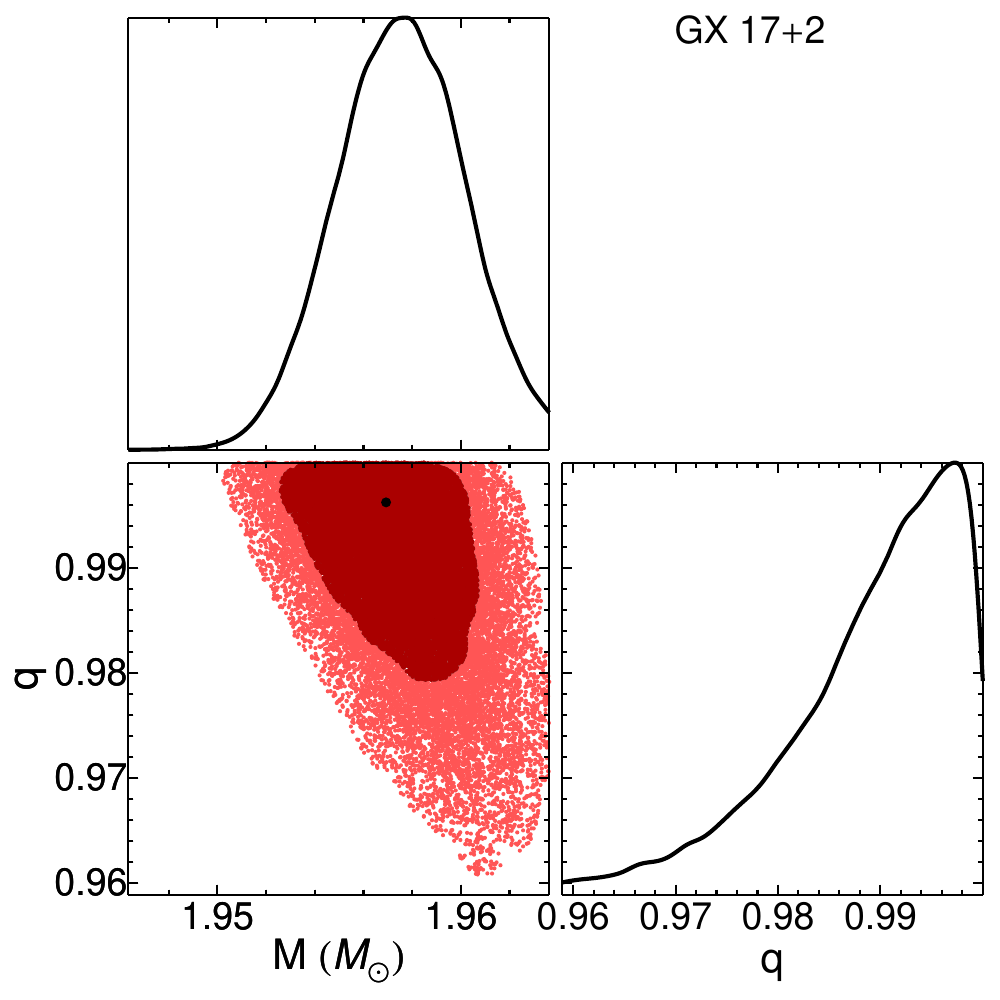}\hfill
\includegraphics[width=0.31\hsize,clip]{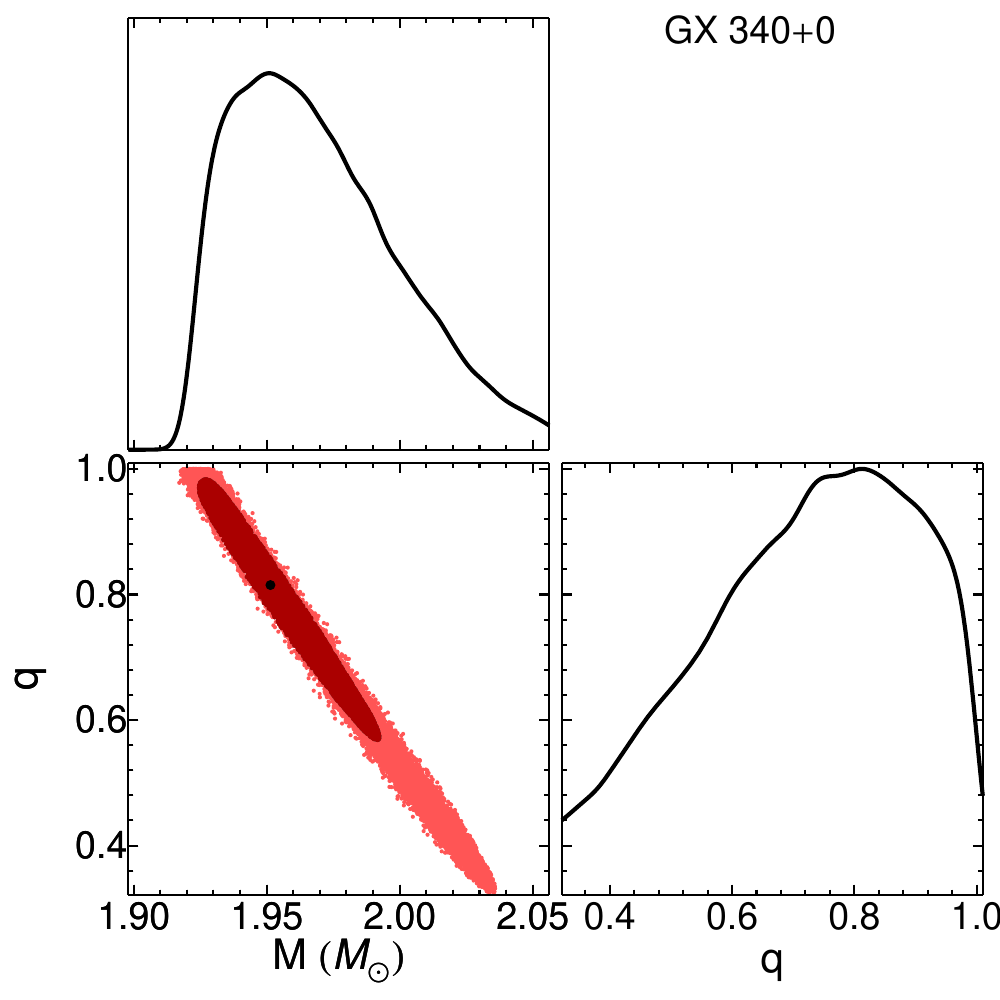}\hfill}

{\hfill
\includegraphics[width=0.31\hsize,clip]{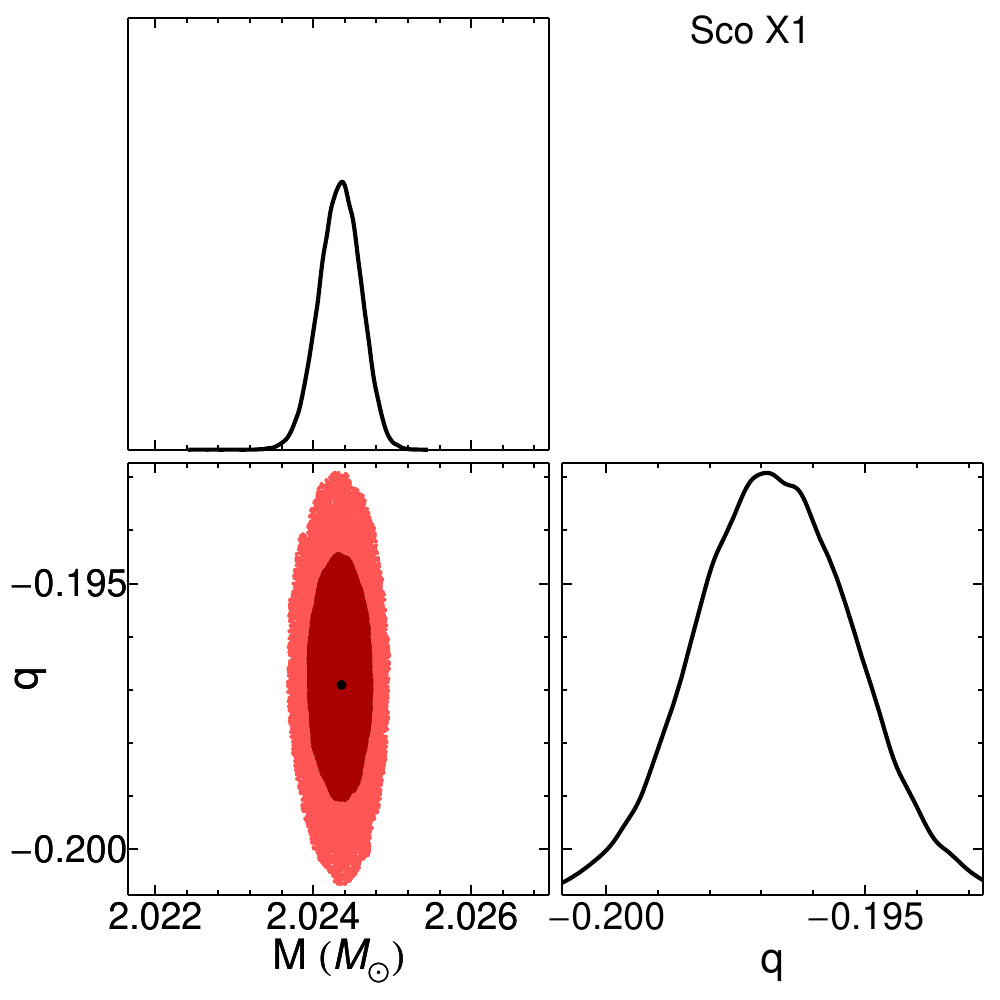}\hfill
\includegraphics[width=0.31\hsize,clip]{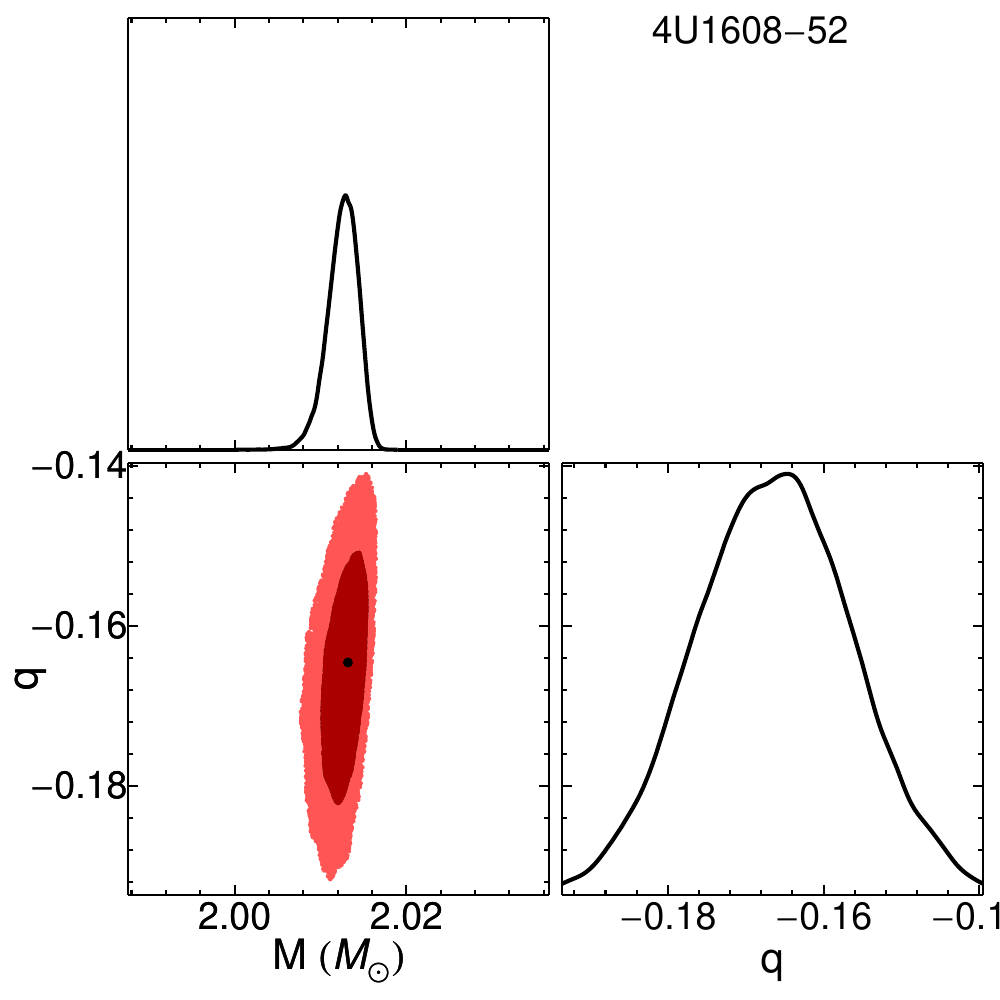}
\hfill}

{\hfill
\includegraphics[width=0.31\hsize,clip]{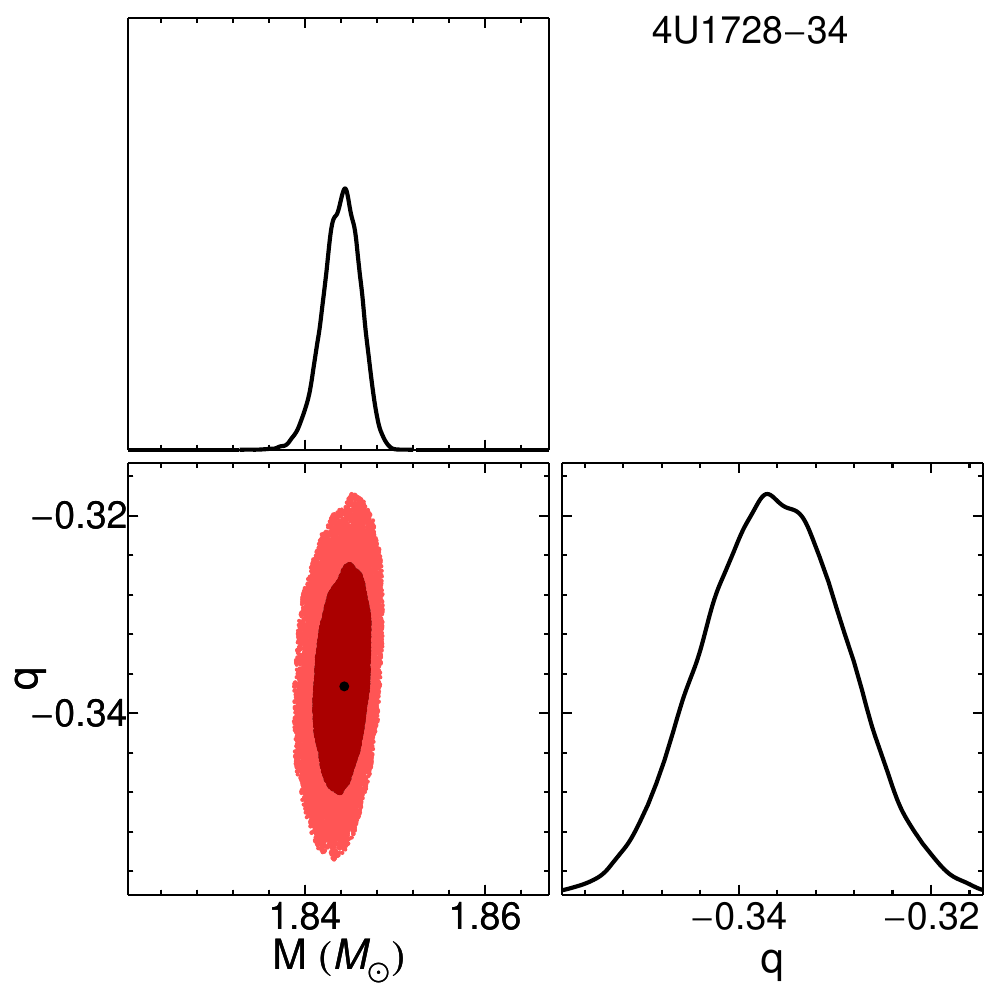}\hfill
\includegraphics[width=0.31\hsize,clip]{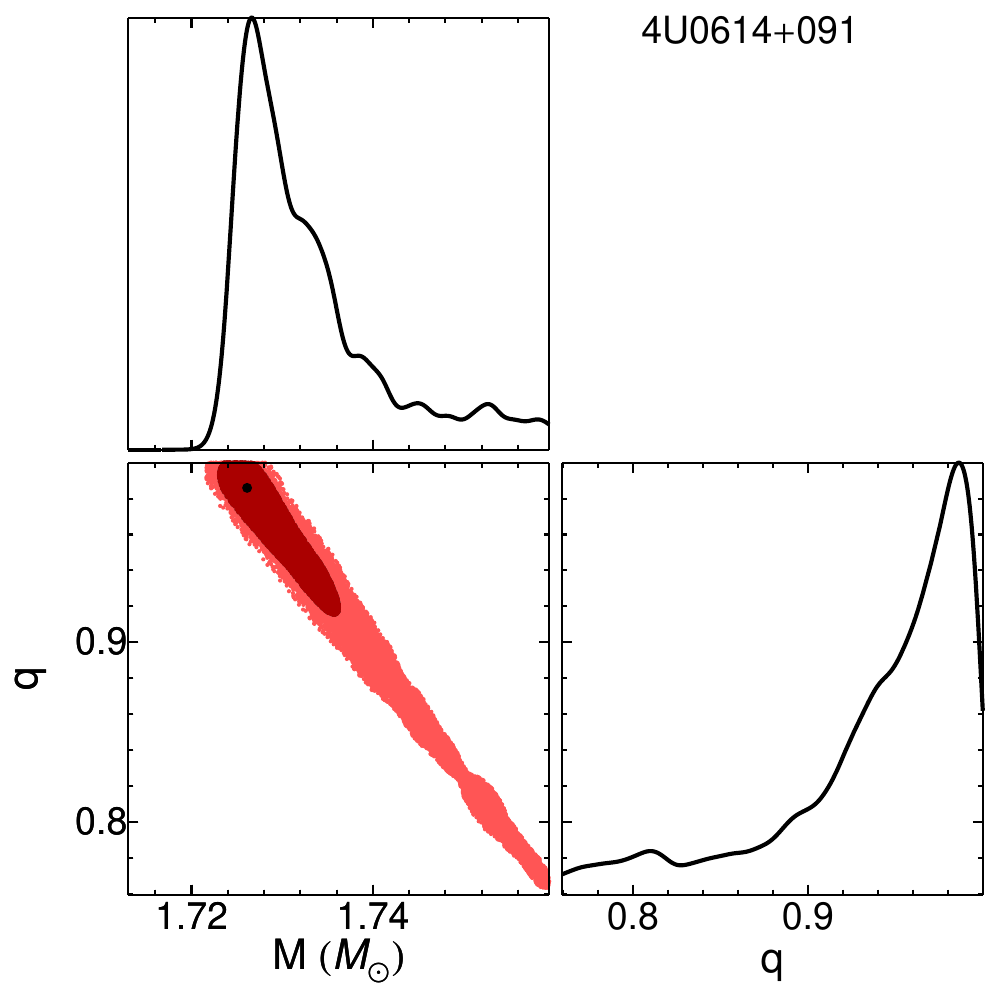}\hfill}
\caption{The ZV metric contours plots of the best-fit parameters (black dots) and the associated 1--$\sigma$ (dark) and 2--$\sigma$ (light) confidence regions of the sources listed in Tab. \ref{tab:results_MCMC}.} 
\label{fig:contoursZV}
\end{figure*}

\begin{figure*}
{\hfill
\includegraphics[width=0.31\hsize,clip]{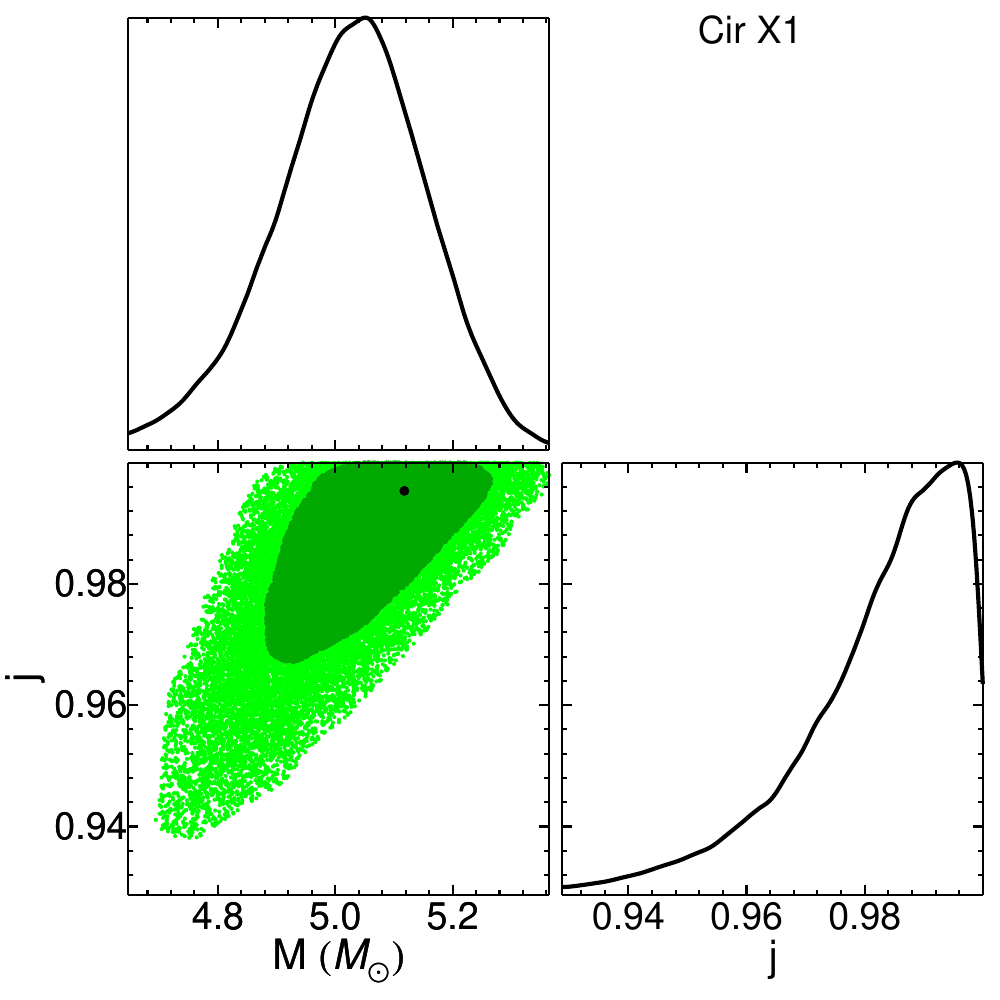}\hfill
\includegraphics[width=0.31\hsize,clip]{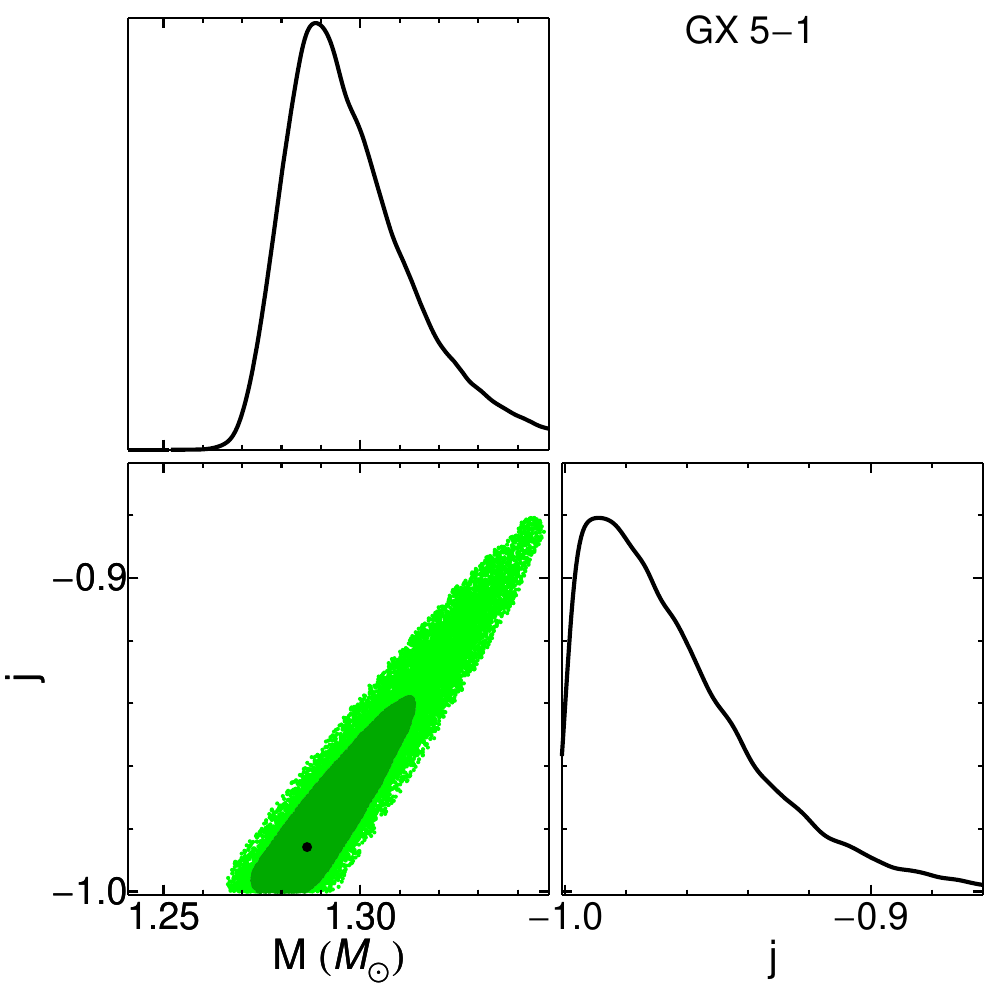}
\hfill}

{\hfill
\includegraphics[width=0.31\hsize,clip]{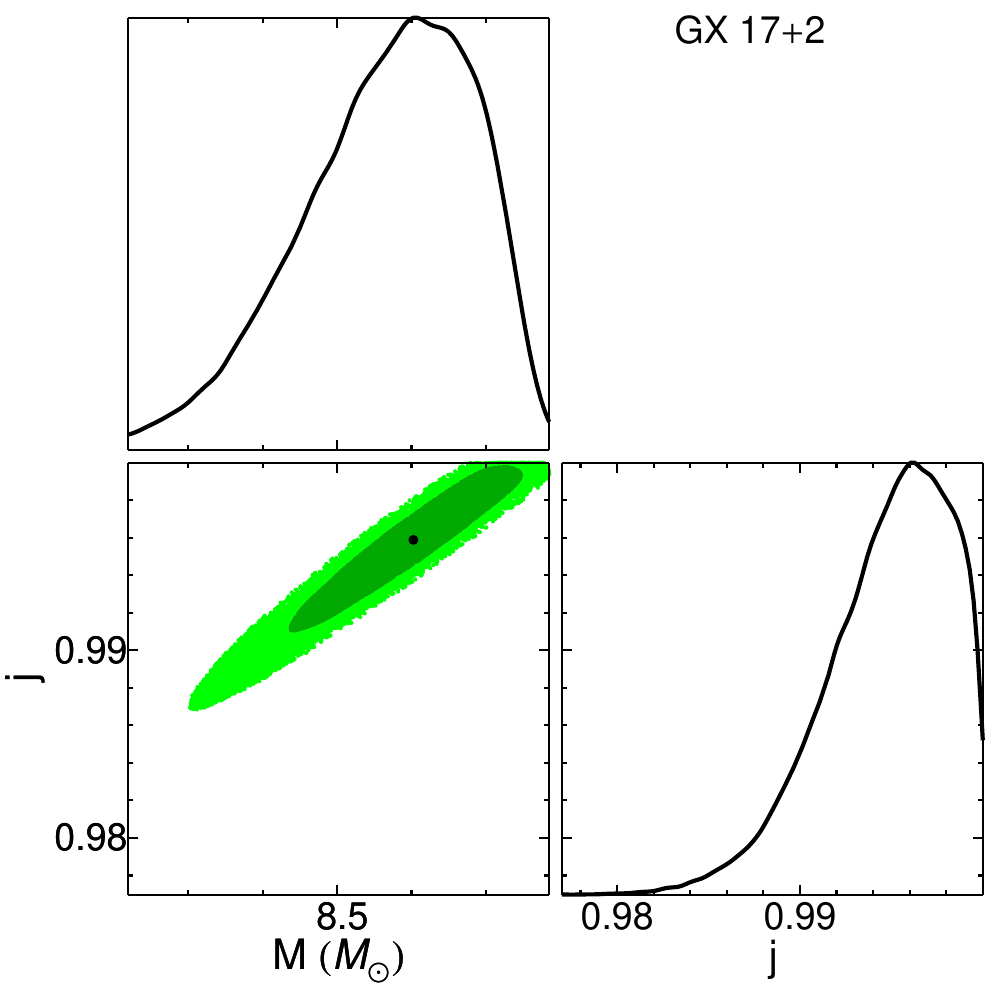}\hfill
\includegraphics[width=0.31\hsize,clip]{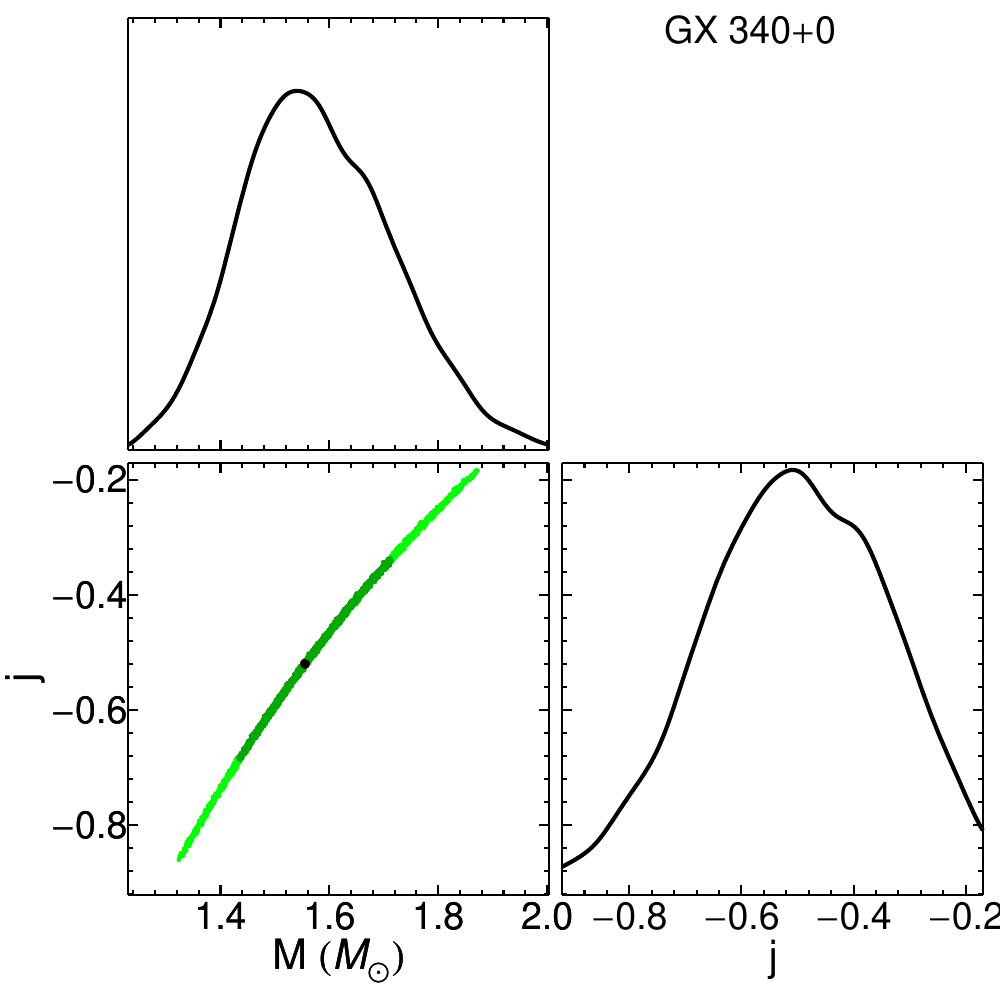}\hfill}

{\hfill
\includegraphics[width=0.31\hsize,clip]{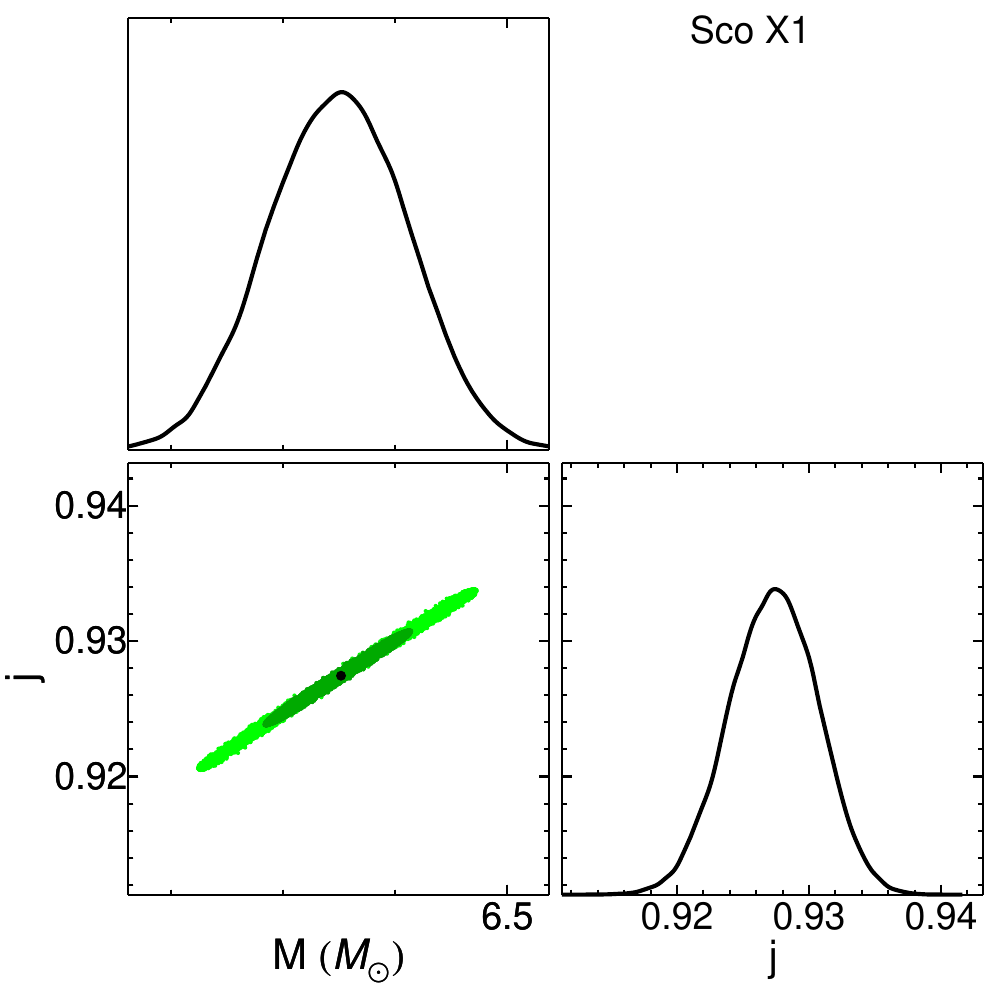}\hfill
\includegraphics[width=0.31\hsize,clip]{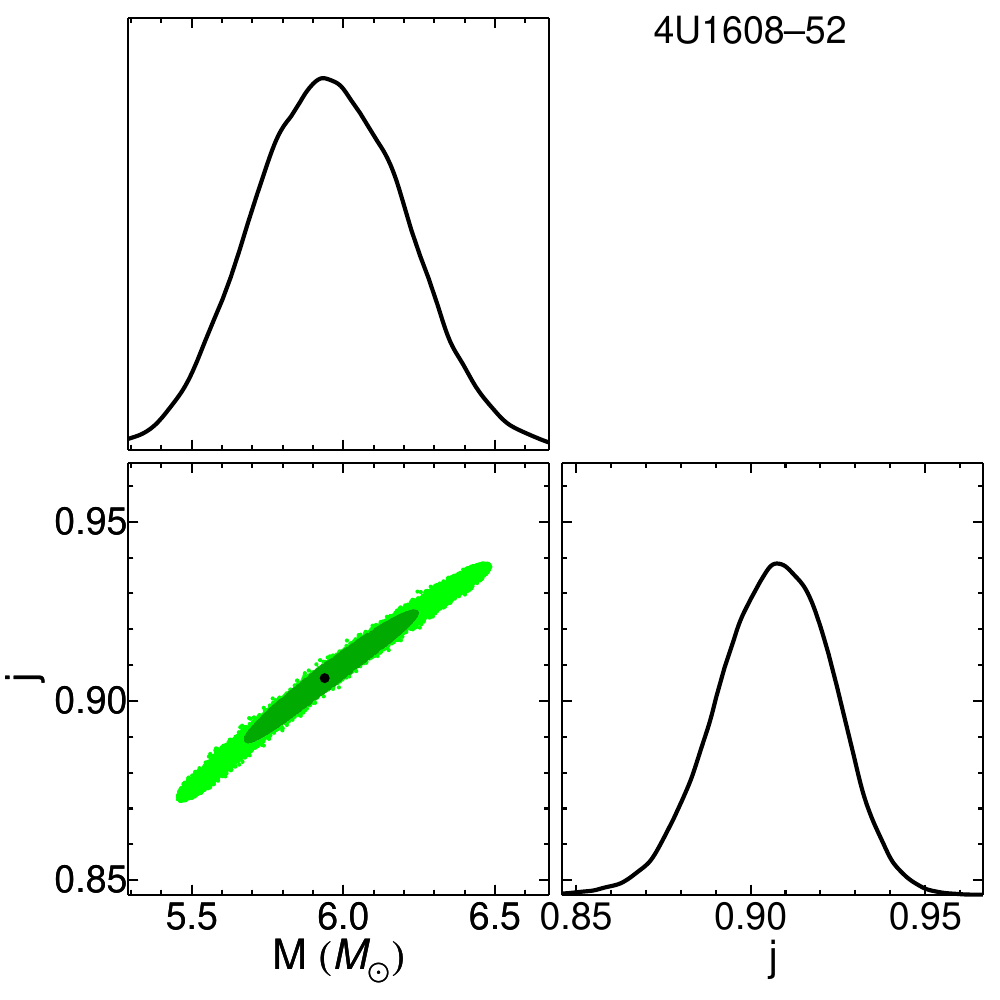}
\hfill}

{\hfill
\includegraphics[width=0.31\hsize,clip]{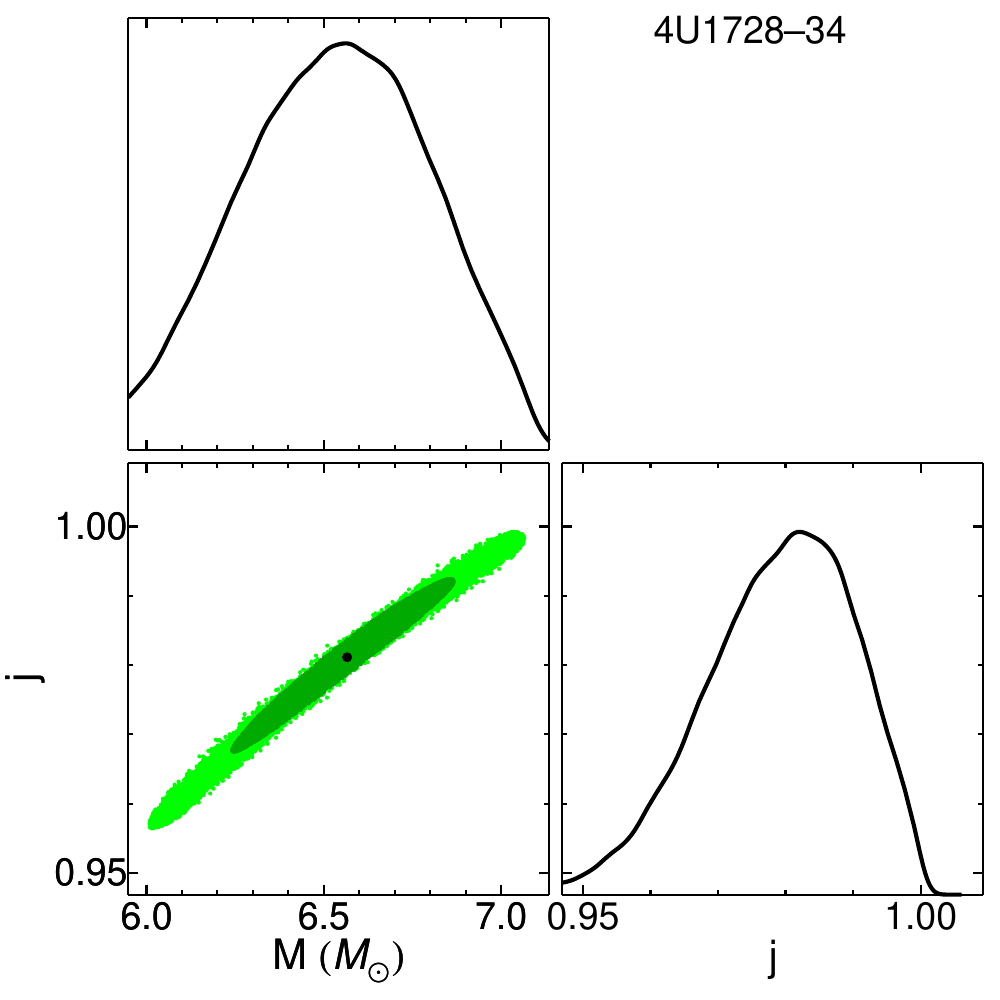}\hfill
\includegraphics[width=0.31\hsize,clip]{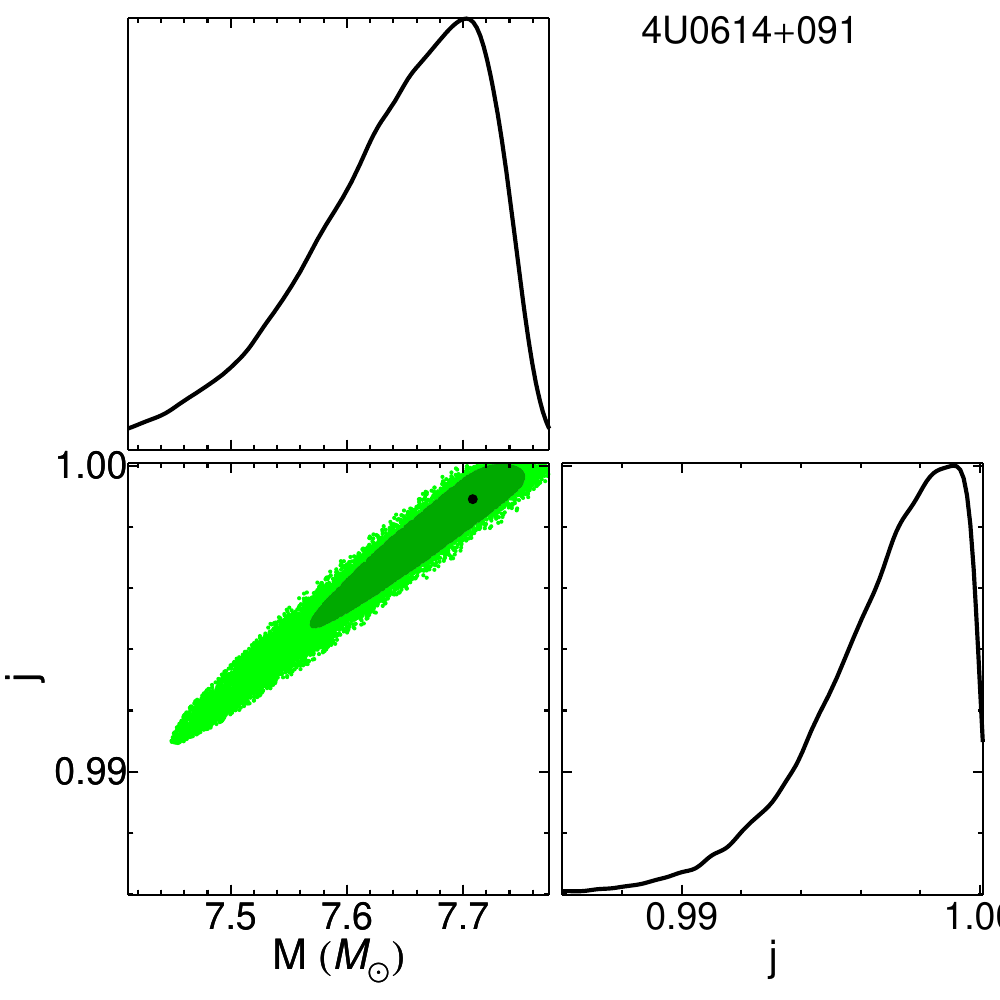}\hfill}
\caption{The same as in Fig.~\ref{fig:contoursZV} but for the Kerr metric.} 
\label{fig:contoursK}
\end{figure*}

\begin{figure*}
{\hfill
\includegraphics[width=0.31\hsize,clip]{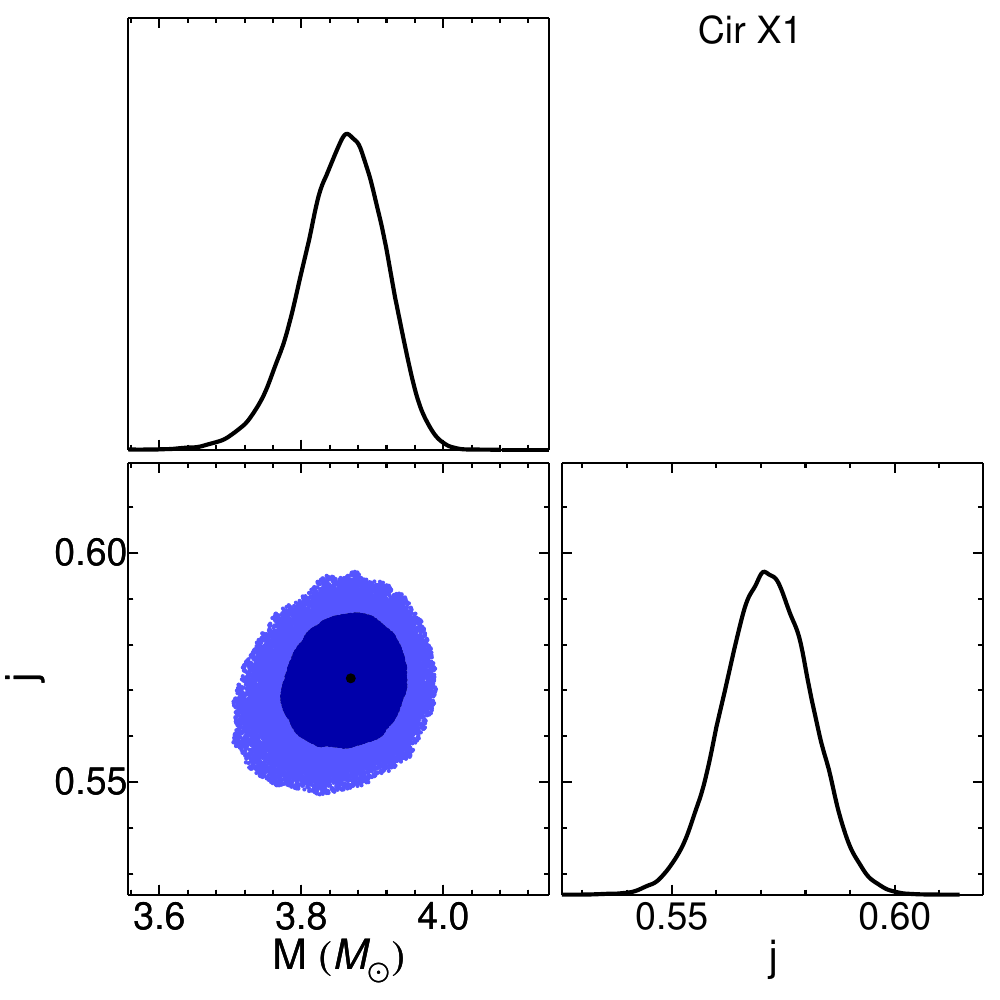}\hfill
\includegraphics[width=0.31\hsize,clip]{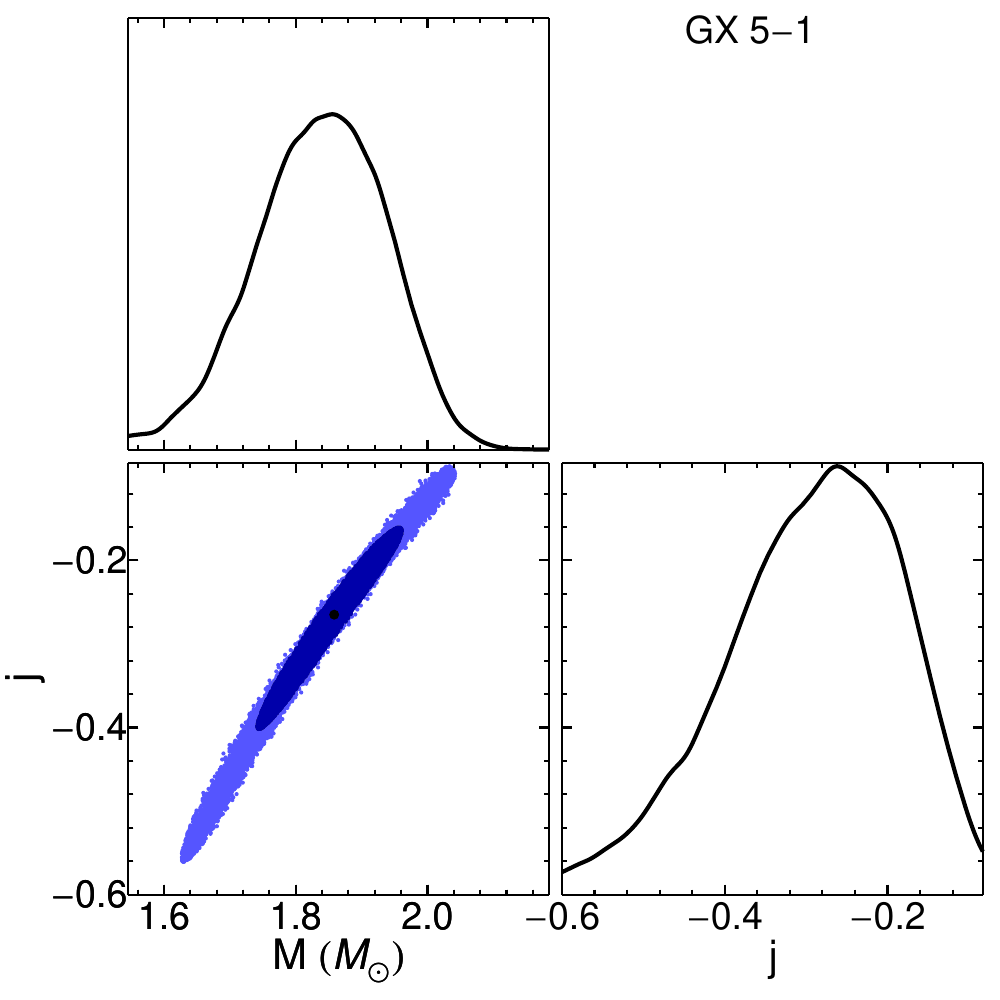}
\hfill}

{\hfill
\includegraphics[width=0.31\hsize,clip]{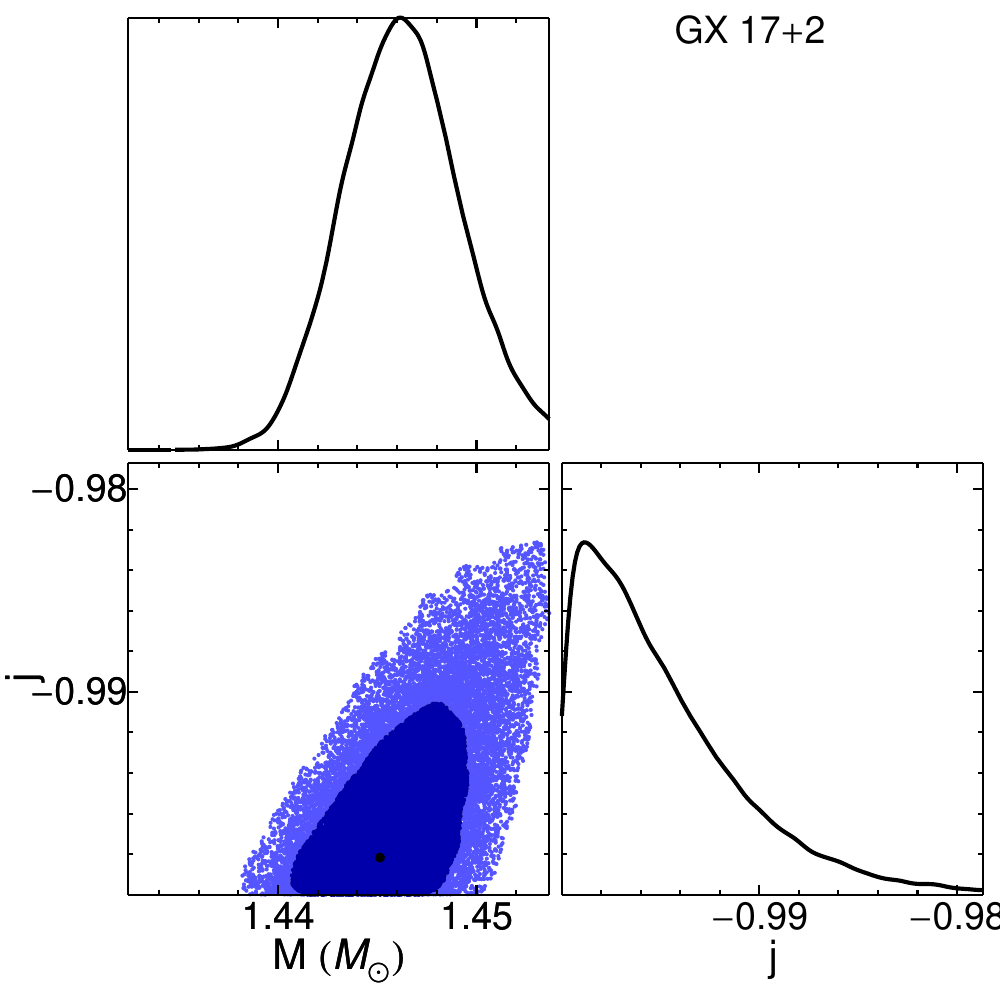}\hfill
\includegraphics[width=0.31\hsize,clip]{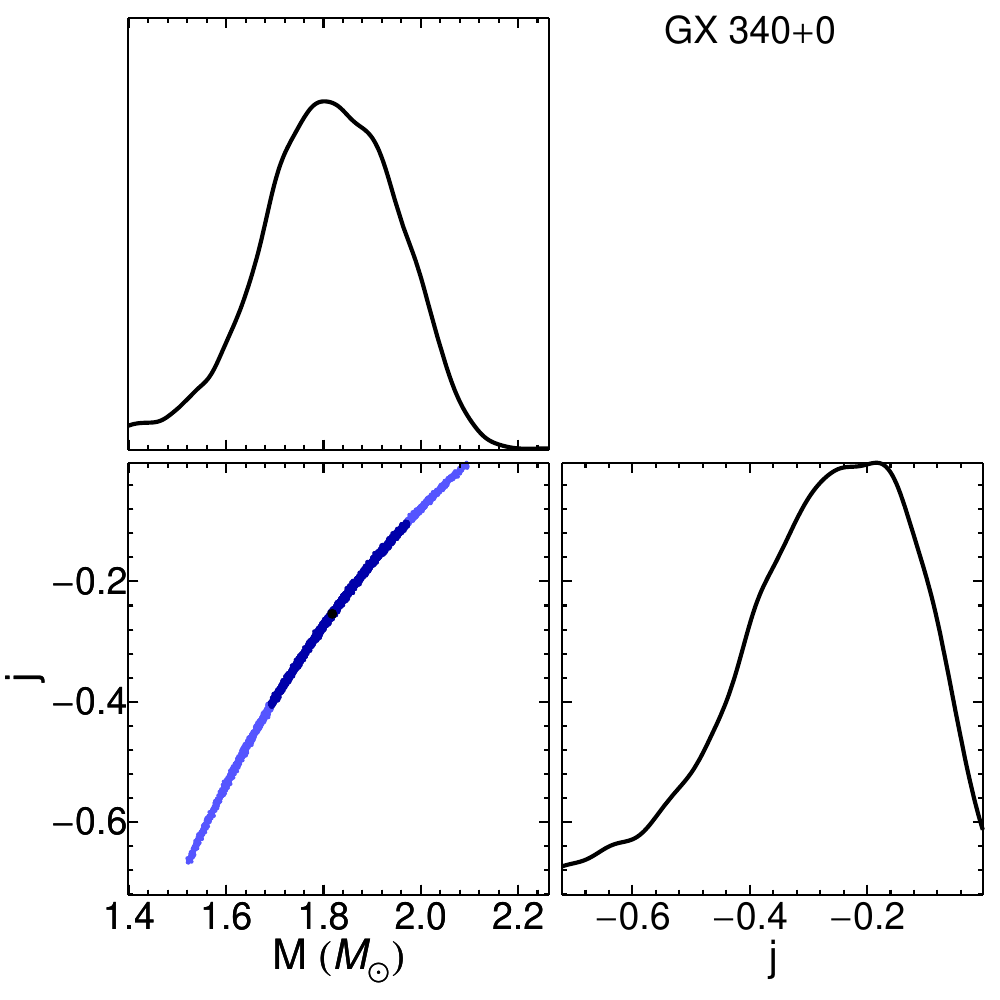}\hfill}

{\hfill
\includegraphics[width=0.31\hsize,clip]{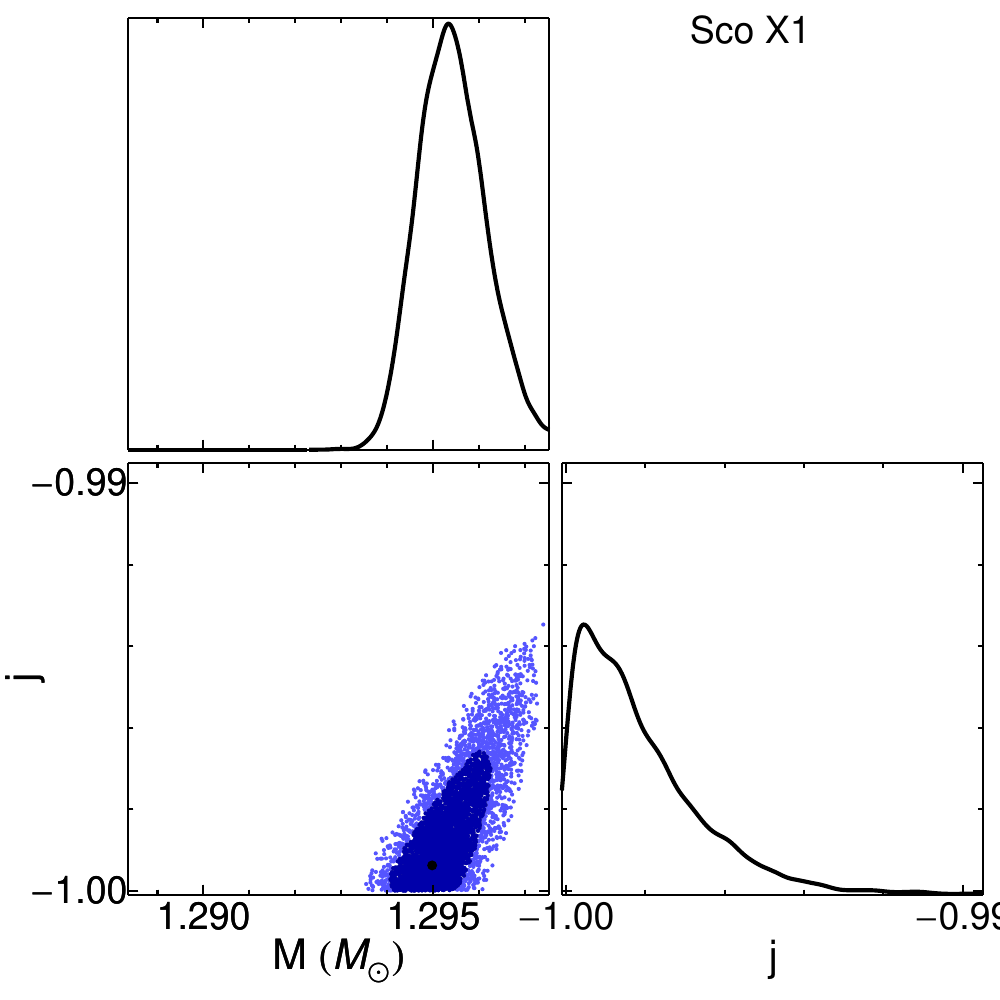}\hfill
\includegraphics[width=0.31\hsize,clip]{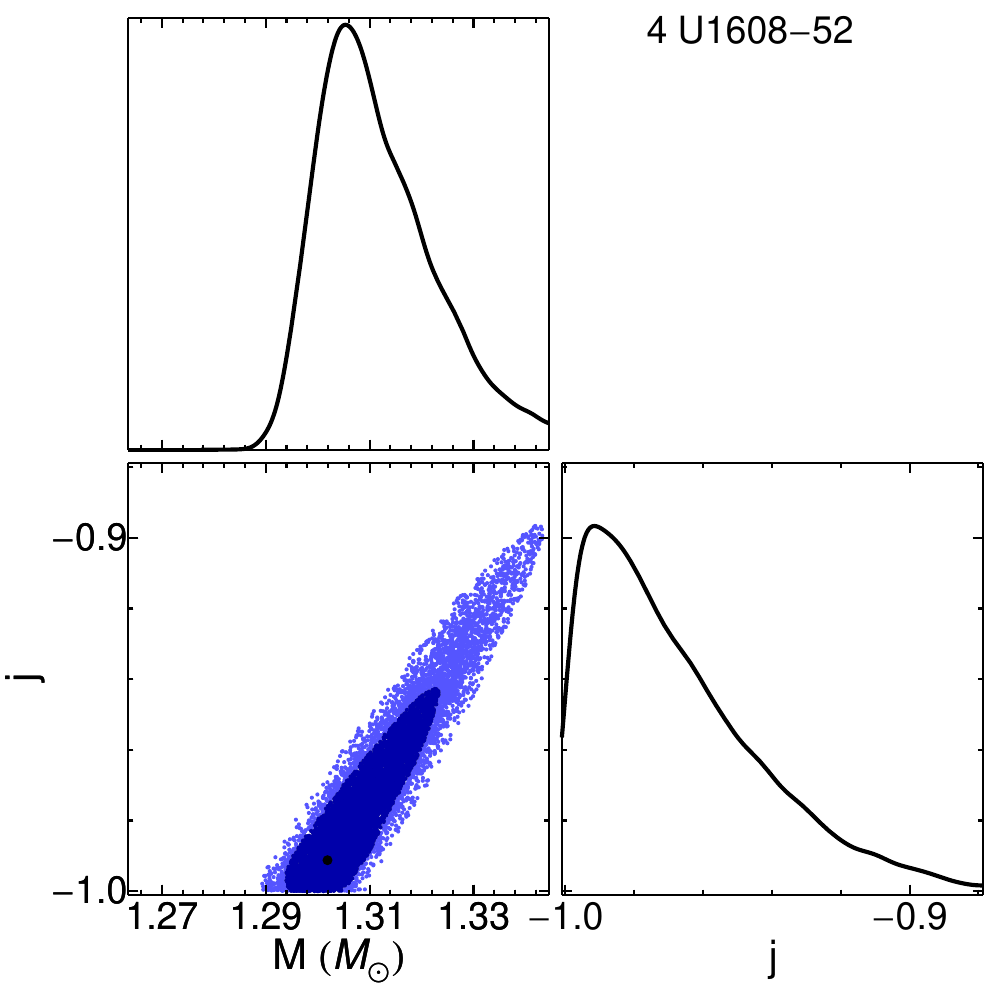}
\hfill}

{\hfill
\includegraphics[width=0.31\hsize,clip]{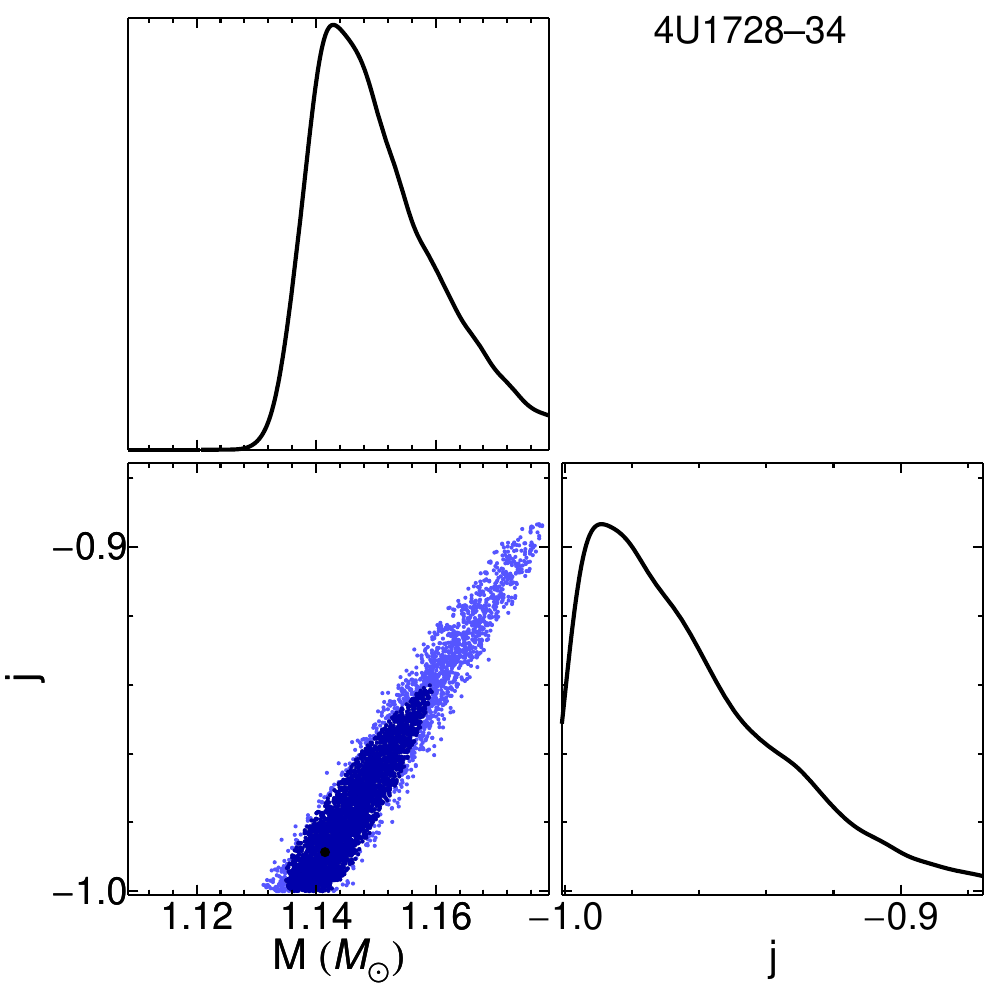}\hfill
\includegraphics[width=0.31\hsize,clip]{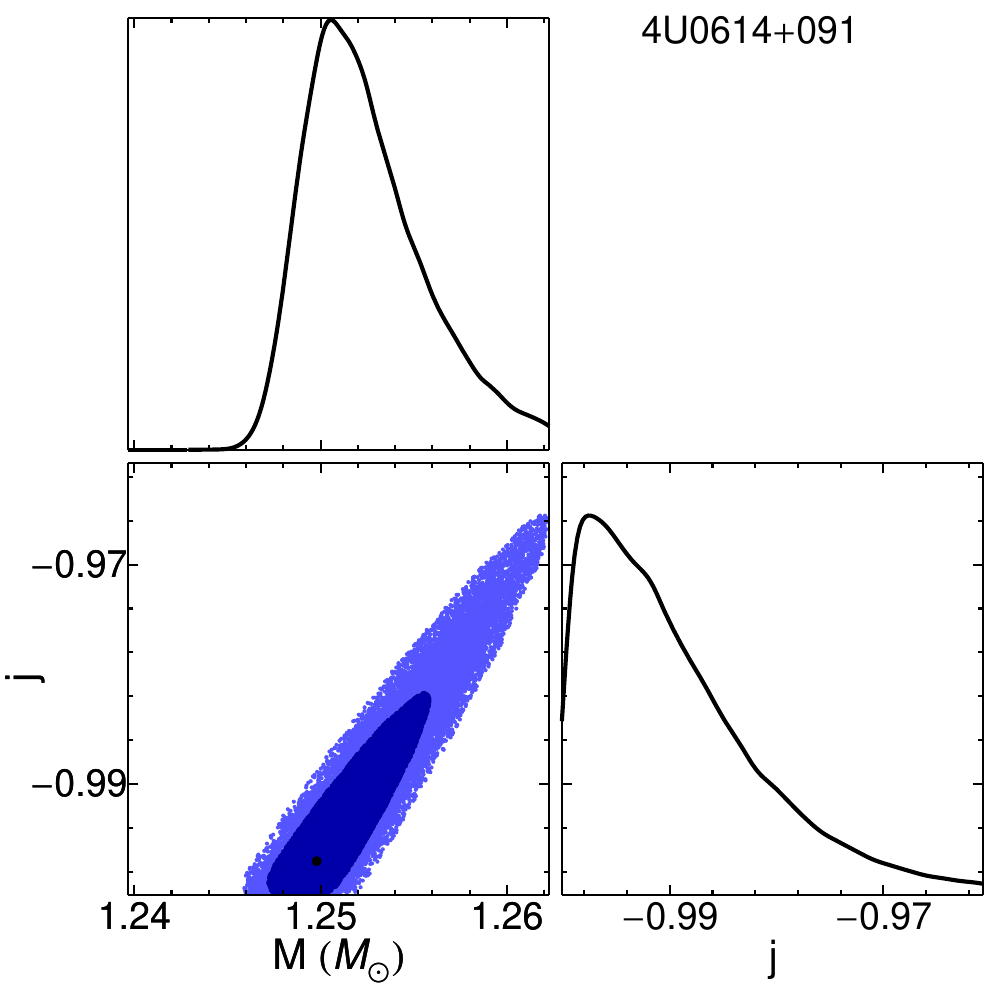}\hfill}
\caption{The same as in Fig.~\ref{fig:contoursZV} but for the LT metric.} 
\label{fig:contoursLT}
\end{figure*}

Here we demonstrate the contour plots obtained from our fits for different solutions of the field equations. Figs.~\ref{fig:contoursZV},\ref{fig:contoursK} and  \ref{fig:contoursLT} present our analyses for the ZV, Kerr and LT metrics, respectively. 


\bibliographystyle{mnras}
\bibliography{0refs} 

\end{document}